\documentclass[aps,groupedaddress,a4paper,floatfix,twocolumn,footinbib,showpacs]{revtex4}
\usepackage{pslatex}
\usepackage{amsfonts}
\usepackage{amssymb}
\usepackage{mathrsfs}
\usepackage[mathscr]{euscript}
\usepackage{graphicx}
\usepackage{ifthen}
\usepackage{fancyhdr}
\newboolean{BoolLong}
\newboolean{BoolLabels}
\newboolean{BoolSHG}
\newboolean{BoolDPA}
\newboolean{BoolDPD}
\newboolean{BoolNPA}
\newboolean{BoolNPD}

\newboolean{BoolOPOall}
\newboolean{BoolOPOsync}
\newboolean{BoolOPOfixed}

\newboolean{BoolExternalBib}
\setboolean{BoolExternalBib}{true}

\setboolean{BoolLong}{true}

\setboolean{BoolLabels}{false}

\ifthenelse{\boolean{BoolLong}}
{
\setboolean{BoolSHG}{true}
\setboolean{BoolDPA}{true}
\setboolean{BoolDPD}{true}
\setboolean{BoolNPA}{true}
\setboolean{BoolNPD}{true}
\setboolean{BoolOPOall}{true}
\setboolean{BoolOPOsync}{true}
\setboolean{BoolOPOfixed}{true}
}
{
\setboolean{BoolSHG}{false}
\setboolean{BoolDPA}{false}
\setboolean{BoolDPD}{false}
\setboolean{BoolNPA}{true}
\setboolean{BoolNPD}{true}
\setboolean{BoolOPOall}{false}
\setboolean{BoolOPOsync}{false}
\setboolean{BoolOPOfixed}{false}
}

\def\overstrike#1#2{{\setbox0\hbox{$#2$}\hbox to \wd0{\hss
    $#1$\hss}\kern-\wd0\box0}}

\begin{document}

\title{Few Cycle Pulse Propagation}
\author{P. Kinsler}
\affiliation{
  Department of Physics, Imperial College,
  Prince Consort Road,
  London SW7 2BW, 
  United Kingdom.
}
\author{G.H.C. New}
\affiliation{
  Department of Physics, Imperial College,
  Prince Consort Road,
  London SW7 2BW, 
  United Kingdom.
}

\lhead{
\includegraphics[height=5mm,angle=0]{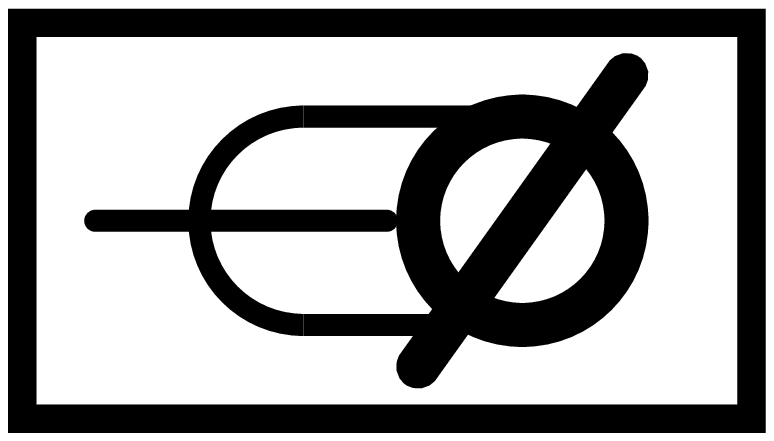}~~
FCOPO(long)}
\chead{~}
\rhead{
\href{mailto:Dr.Paul.Kinsler@physics.org}{Dr.Paul.Kinsler@physics.org}\\
\href{http://www.kinsler.org/physics/}{http://www.kinsler.org/physics/}
}

\date{\today}

\begin{abstract}

We present a comprehensive framework for treating the nonlinear interaction of
few-cycle pulses using an envelope description that goes beyond the traditional
SVEA method.  This is applied to 
a range of simulations that demonstrate how the effect of
a $\chi^{(2)}$ nonlinearity differs between the many-cycle and few-cycle
cases.  Our approach, which includes diffraction, dispersion, multiple fields,
and a wide range of nonlinearities, builds upon the work of Brabec and
Krausz\cite{Brabec-K-1997prl} and Porras\cite{Porras-1999pra}.  No
approximations are made until the final stage when a particular problem is
considered.

 The original version (v1) of this arXiv paper is close to the 
published Phys.Rev.A. version, and much smaller in size.

\end{abstract}


\maketitle
\thispagestyle{fancy}

{\em
NOTE: this is a longer version of a paper published as 
Phys. Rev. A {\bf 67}, 023813 (2003).
 It therefore contains the same text and graphics, 
but has additional remarks and many extra figures, most notably more
complete sets of simulation results.   This has led to some 
repetitions in the text, and also some of the extra figures have 
artifacts: for example, the envelope phase graphs when the 
envelope has a small magnitude, and some of the figures 
showing the phase drift have spikes where a stable value could 
not be found.  Further, some figures pass without much comment.
}

\begin{section}{Introduction}\label{intro}

The analysis of optical pulse propagation traditionally involves describing a
pulse in terms of a complex field envelope, while neglecting the underlying
rapid oscillations at its carrier frequency. The resulting ``slowly varying
envelope approximation'' (SVEA)  (see e.g. \cite{Shen-PNLO}), which reduces
second order differential equations to first order, is valid when the envelope
encompasses many cycles of the optical field and varies slowly.
The alternative approach is to solve Maxwell's equations
numerically (see e.g.  \cite{Brabec-K-1997prl,Tarasishin-MZ-2001oc}), which is
more general but involves greater computational effort, and lacks the 
intuitive picture of a pulse ``envelope''.

For example, optical parametric oscillators (OPOs) based on
aperiodically-poled lithium niobate (APPLN) have generated 53 fs idler pulses
at 3$\mu$m that are nearly transform limited, and contain only five optical
cycles \cite{Beddard-ERS-2000ol}; laser pulses with less than three optical
cycles have been generated in other contexts \cite{Baltuska-WPW-1997ol}.  Under
these circumstances, the validity of the slowly-varying envelope approximation
is clearly open to question.

Brabec and Krausz \cite{Brabec-K-1997prl} derived corrections to the SVEA,
which they included in their ``slowly evolving wave approximation'' (SEWA). 
This enabled the few-cycle regime to be modelled with improved accuracy, and
the SEWA has subsequently been applied in different situations, including
ultrashort IR laser pulses in fused silica 
\cite{Ranka-G-1998ol,Tzortzakis-SFPMCB-2001prl},
the filamentation of ultra-short laser pulses in air
\cite{Akozbek-SBC-2001oc}, and even in microstructured optical fibres
\cite{Gaeta-2002ol}.  Later, Porras \cite{Porras-1999pra} proposed a slightly
different ``slowly evolving envelope approximation'' (SEEA) that included
corrections for the transverse behavior of the field.

Here we use a field envelope approach to simulate the propagation of
ultrashort pulses in a $\chi^{(2)} $medium.  The novelty is that we (a)
derive a more general form than that of Brabec and Krausz, called the
``generalised few-cycle envelope approximation''(GFEA); and (b) apply it to
both optical (non-degenerate) parametric amplification (NPA), and the optical
parametric oscillator (OPO). 
A shorter version of this paper has been published in Phys. Rev. A 
\cite{Kinsler-N-2003pra}.
The only previous multiple field application
of this kind of result was for four wave mixing
\cite{GallagherFaeder-J-2000pra}.  

We compare the SEWA/SEEA equations to our own (section \ref{theory}), and
explain the differences and subsequent adjustments to the necessary
approximations.  This theory enables us to rigorously study what combination
of approximations affords the most efficient method for treating a given
nonlinear interaction involving few-cycle pulses.  Next (section \ref{chi2}) 
we discuss the 
$\chi^{(2)}$ nonlinearity and a scaling
scheme designed to reveal the few-cycle effects.  Then
we compare the SVEA predictions to the few cycle GFEA theory using idealised
situations  (section \ref{idealchi2}) and more realistic OPO models (section
\ref{opo}).  Finally, we present our conclusions (section \ref{conclude}).

\end{section}


\begin{section}{Theory}\label{theory}

This section contains a summary of a complete rederivation
\cite{Kinsler-FCPP} of a Brabec \& Krausz style theory, which yields an
evolution equation for an envelope description of pulse
propagation in the few-cycle regime.  Our result is more complicated 
than the SEWA equation \cite{Brabec-K-1997prl}, but reduces to it
in the appropriate limits; it also explains the slight differences
between their result and that of Porras \cite{Porras-1999pra}.

Following Brabec-Krausz, we consider the case of small transverse
inhomogeneities of the polarization, and so start with the three dimensional
wave equation

\begin{eqnarray}
&&
  \left( \partial_z^2 + \nabla_\bot^2 \right) E(\vec{r},t)
- \frac{1}{c^2}
  \partial_t^2
  \int_{-\infty}^t dt' \epsilon(t-t') E(\vec{r},t')
\nonumber 
\\
&&=
  \frac{4\pi}{c^2}
  \partial_t^2
  P_{nl}(\vec{r},t)
.
\label{eqn-3DWE-time}
\end{eqnarray}

Here $\nabla_\bot^2$ is the transverse Laplace operator,
$\partial_\alpha$ is shorthand notation for $\partial/\partial
\alpha$, $\epsilon(t) = (2\pi)^{-1} \int_{-\infty}^\infty d\omega
\tilde{\epsilon}(\omega) e^{\imath \omega t}$, $\tilde{\epsilon}(\omega) = 1 +
4\pi \chi(\omega)$, and $\chi(\omega)$ is the linear electric susceptibility. 
The electric field $E$ propagates along the $z$ direction. Both $E$ and the
nonlinear polarization $P_{nl}$ are polarized parallel to the $x$ axis.

We can transform eqn. (\ref{eqn-3DWE-time}) into frequency space in order to
expand $\tilde{\epsilon}(\omega)$ in powers of $\omega$, thus enabling us to
treat the material parameters as a power series which we can truncate to an
approriate order.  However for simplicity it is better to expand $k$ about a
suitable $\omega_0$ instead.  Using $\tilde{\epsilon}(\omega) = c^2
k(\omega)^2 / \omega^2$, it follows that

\begin{eqnarray}
&&
 k(\omega)=\sum_{n=0}^\infty 
             \frac{\gamma_n \left( \omega - \omega_0 \right)^n}
                  {n!}
;
\nonumber
\\
&&
  \gamma_n  = \left. \partial_\omega^n k(\omega) \right|_{\omega_0}
            = \beta_n + \imath \alpha_n
; 
  \beta_n, \alpha_n \in \mathbb{R}
.
\label{eqn-expansion-coeffs}
\end{eqnarray}

We can now write the frequency space version of eqn. (\ref{eqn-3DWE-time}) as 

\begin{eqnarray}
&&
  \left( \partial_z^2 + \nabla_\bot^2 \right)
    E(\vec{r},t)
 +
     \left[
       \sum_{n=0}^\infty 
       \frac{\imath^n \gamma_n 
                      \left( \partial_t +\imath \omega_0
                      \right)^n
            }
            {n!}
     \right]^2
        E(\vec{r},t) 
\nonumber
\\
&&=
  \frac{4\pi}{c^2}
  \partial_t^2
    P_{nl}(\vec{r},t) 
\label{eqn-3DWE-w-expand}
.
\end{eqnarray}

We introduce an envelope and carrier form for the field in the 
usual way, using $\vec{r} \equiv \left( \vec{r}_\bot, z \right)$, 
so that

\begin{eqnarray}
  E(\vec{r},t) 
=&& 
  A(\vec{r}_\bot,z,t) 
  e^{\imath \Xi}
+
  A^*(\vec{r}_\bot,z,t) 
  e^{-\imath \Xi}
\label{eqn-envelopecarrier}
,
\end{eqnarray}

\noindent 
and similarly $P_{nl}(\vec{r},t) = B(\vec{r}_\bot,z,t ; A) e^{\imath \Xi}
+ B^*(\vec{r}_\bot,z,t ; A) e^{-\imath \Xi}$.  The symbol $\Xi=\beta_0 z -
\omega_0 t + \psi_0$ is introduced as a convenient shorthand
for the argument of the exponential.  
With these envelope-carrier substitutions, the equation of motion 
can be written as

\begin{eqnarray}
&&
  \left( \left[\imath \beta_0 + \partial_z \right]^2 + \nabla_\bot^2 \right)
    A(\vec{r}_\bot,z,t)
\nonumber
\\
&+&
     \left[
       \sum_{n=0}^\infty 
          \frac{\gamma_n \omega_0^n }{n!}
          \left( \frac{\imath}{\omega_0} \partial_t \right)^n
     \right]^2
        A(\vec{r}_\bot,z,t) 
\nonumber
\\
&=&
 -
  \frac{4 \pi \omega_0^2}{c^2}
  \left(1 + \frac{\imath}{\omega_0}\partial_t \right)^2
    B(\vec{r}_\bot,z,t ; A) 
.
\label{nearlyBKeqn2}
\end{eqnarray}

\noindent 
Eqn. (\ref{nearlyBKeqn2}) has the opposite sign on the RHS 
to Brabec \& Krausz's eqn. (2), but agreement is recovered later in
eqn. (\ref{exact-BKP}).

As is usual, we introduce co-moving variables 

\begin{eqnarray}
\tau&=&\omega_0 \left(t-\beta_1 z\right)
, ~~~~
  \partial_t 
\equiv 
  \omega_0 \partial_\tau
,
\\
\xi&=&\beta_0 z 
, ~~~~
  \partial_z 
\equiv 
  \beta_0 \partial_\xi - \omega_0 \beta_1 \partial_\tau
\label{coords-coscaled}
,
\end{eqnarray}

\noindent 
and eqn. (\ref{nearlyBKeqn2}) now becomes 

\begin{eqnarray}
&&
  \left\{ 
    \left(  \imath \beta_0 
               + \beta_0 \partial_\xi - \omega_0 \beta_1 \partial_\tau
    \right)^2 
  + \nabla_\bot^2 
  +
     \left[
       \sum_{n=0}^\infty 
          \frac{\gamma_n \omega_0^n }{n!}
          \left(\imath \partial_\tau \right)^n
     \right]^2
  \right\}
\nonumber
\\
&&\times
        A(\vec{r}_\bot,\xi,\tau) 
 +
  \frac{4 \pi \omega_0^2}{c^2}
  \left(1 + \imath \partial_\tau \right)^2
    B(\vec{r}_\bot,\xi,\tau ; A) 
=0
.
\label{start9669886}
\end{eqnarray}

For convenience we also introduce the dimensionless ratio of phase and group
velocities $\sigma = \omega_0 \beta_1 / \beta_0 = 
\left( \omega_0 / \beta_0 \right) /
(1/\beta_1) = v_f / v_g $, and use the fact that the refractive index at
$\omega_0$ is $n_0= c \beta_0 / \omega_0$.  We also define a dispersion term
$\hat{D}$ in a similar way to Brabec-Krausz, but instead use a scaled
(dimensionless) version $\hat{D}' = \left(\omega_0/\beta_0\right) \hat{D}$
in following equations so that 

\begin{eqnarray}
\hat{D}'
&=&         
  \frac{ \omega_0}{\beta_0}
  \left[
    \imath \alpha_1
    \left(\imath \partial_\tau \right)
   +
    \sum_{n=2}^\infty 
          \frac{\gamma_n \omega_0^{n-1} }{n!}
          \left(\imath \partial_\tau \right)^n
  \right]
. ~~~~
\end{eqnarray}

Hence we get 

\begin{eqnarray}
0
&=&
  \left\{ 
      \left(  \partial_\xi 
            - \sigma \partial_\tau
      \right)
    + 
      \frac{1}{2\imath}
      \left(  \partial_\xi 
            - \sigma \partial_\tau
      \right)^2 
    + \frac{1}{2 \imath \beta_0^2}
      \nabla_\bot^2 
\right.
\nonumber
\\
&&    -
      \left[
        \imath
          \sigma
          \left(\imath \partial_\tau \right)
        -
          \frac{\imath \alpha_0}{\beta_0}
        + 
          \imath
          \hat{D}'
      \right]
\nonumber
\\
&&    +
\left.
      \frac{\imath}{2}
      \left[
        \imath
          \sigma
          \left(\imath \partial_\tau \right)
        -
          \frac{\imath \alpha_0}{\beta_0}
        + 
          \imath
          \hat{D}'
      \right]^2
  \right\}
        A(\vec{r}_\bot,\xi,\tau) 
\nonumber
\\
&& +
  \frac{2 \pi }{\imath n_0^2}
  \left(1 + \imath \partial_\tau \right)^2
    B(\vec{r}_\bot,\xi,\tau ; A) 
.
\end{eqnarray}

This form can be rearranged without approximation to

\begin{eqnarray}
&&\partial_\xi
  A(\vec{r}_\bot,\xi,\tau) 
\nonumber
\\
&=&
    \left( 
    - \frac{ \alpha_0}{\beta_0}
    + \imath  \hat{D}'
    \right)
  A(\vec{r}_\bot,\xi,\tau) 
  + 
    \frac{\left( \imath / 2 \beta_0^2 \right) \nabla_\bot^2}
         { \left( 1 + \imath \sigma \partial_\tau \right)}
    A(\vec{r}_\bot,\xi,\tau) 
\nonumber
\\
&+&
    \frac{2 \imath \pi }{n_0^2}
    \frac{\left(1 + \imath \partial_\tau \right)^2}
         {\left( 1 + \imath \sigma \partial_\tau \right)}
    B(\vec{r}_\bot,\xi,\tau ; A)
+
    \frac{ T_{R} }
         { 1 + \imath \sigma \partial_\tau }
,
\label{exact-BKP}
\end{eqnarray}

\noindent 
where 

\begin{eqnarray}
T_{R} 
&=& 
  \left[
  -
    \frac{\imath q^2 }{2}
    \partial_\xi^2
  +
    \frac{\imath}{2}
    \left(
      \frac{ \alpha_0}{\beta_0}
    - \imath  \hat{D}'
    \right)^2
  \right]
    A(\vec{r}_\bot,\xi,\tau) 
.
\end{eqnarray}

Eqn. (\ref{exact-BKP}) {\em is exact} -- it contains no more approximations
than our starting point eqn. (\ref{eqn-3DWE-time}) except for the expansion 
of $\epsilon$ in powers of $\omega$.  We recover the full
field $E$ from eqn. (\ref{eqn-envelopecarrier}) by recombining $A$ and knowing
the carrier.  The partial derivatives ($\imath \partial_\tau$) in the
denominators can, if necessary,  be treated by Fourier transforming into the
conjugate frequency space ($\Omega$).  Note that like
$\tau$, $\Omega$ is scaled relative to the carrier frequency.

If we set  $T_{R}=0$, this gives us a generalised few cycle envelope (GFEA)
equation, which contains the SVEA \cite{Shen-PNLO},
SEWA \cite{Brabec-K-1997prl}, and SEEA \cite{Porras-1999pra} within it as
special cases. Of course we cannot just set the $T_{R}$  term to zero without
some justification, and this is discussed below.  

The $\left(2 \imath \pi / n_0^2 \right) \mathcal{K} B$
polarization term from eqn. (\ref{exact-BKP}) has prefactors which 
depend on the time-derivative of the polarization, and these new
terms are what add the effect of finite pulse lengths to the pulse
evolution.  
Note that we can write this polarization term in different forms:

\begin{eqnarray}
\mathcal{K}
&=&
    \frac{\left(1 + \imath \partial_\tau \right)^2}
         {\left( 1 + \imath \sigma \partial_\tau \right)}
\nonumber
\\
&=&
    \left(1 + \imath \sigma \partial_\tau \right)
    \left[
       1
     +
       \left(1-\sigma\right)
       \frac{ 2 \imath \partial_\tau
           + 
           \left(1+\sigma\right)
           \partial_\tau^2
           }
           {\left( 1 + \imath \sigma \partial_\tau \right)^2}
    \right]
\nonumber
\\
&=&
    \left(1 + \imath \partial_\tau \right)
      \left[
        1
      +
        \frac{\imath \left(1 - \sigma \right) \partial_\tau}
             {\left(1 + \imath \sigma \partial_\tau \right)^2}
      \right]
.
\label{exact-BKP-altB}
\end{eqnarray}

With $\sigma=1$, these reduce to the $\mathcal{K}=1 + \imath
\partial_\tau$ SEWA \cite{Brabec-K-1997prl} form.  Similarly, to first order
in $\left(\sigma-1\right)$, one can get the $\mathcal{K}=1 + \imath \sigma
\partial_\tau$ form analogous to
    the SEEA corrections \cite{Porras-1999pra},
    although a strict expansion to first order in $\partial_\tau$ 
\cite{Kinsler-FCPP}
    gives a prefactor of 
      $\left[ 1 + \imath \left(2-\sigma\right) \partial_\tau \right]$.  
Finally, for a SVEA theory,
$\mathcal{K}=1$, since the $\partial_\tau$ terms are assumed to be negligible.

The $T_{R}$ term is negligible if the following conditions hold:  

\begin{description}

\item[{\bf Dispersion: }] terms in $\partial_\tau$ can be neglected  if 
\begin{eqnarray}
  \left|
    \left( 
      \frac{ \omega_0^m \gamma'_m}{\beta_0 m!}
      \Omega^m
    \right)
  \tilde{A}(\vec{r}_\bot,\xi,\Omega)
  \right|
&&\ll
  \left|
     \tilde{A}(\vec{r}_\bot,\xi,\Omega)
  \right|
\label{condition-dispersion}
\end{eqnarray}

\item[{\bf Evolution:}] terms in $\partial_\xi^2$ can be neglected  if 
\begin{eqnarray}
  \left|
    \partial_\xi \tilde{A}(\vec{r}_\bot,\xi,\Omega)
  \right|
\ll
  \left|
    \tilde{A}(\vec{r}_\bot,\xi,\Omega)
  \right|
\label{condition-d2xi-final}
\label{condition-evolution}
,
\end{eqnarray}

\end{description}

\noindent and eqn. (\ref{condition-evolution}) only holds if,
in addition, 

\begin{description}

\item[{\bf Diffraction: }\label{i-diffraction}] terms in $\nabla_\bot^2$ can be neglected  if 
\begin{eqnarray}
    \left(1 + \sigma \Omega \right)
\beta_0^2 w_0^2 
\gg
  1
\label{condition-diffraction}
,
\end{eqnarray}

\item[{\bf Nonlinearity: }\label{i-nonlinearity}] is ``weak'' if
\begin{eqnarray}
  \frac{n_0^2}{2\pi}
    \frac{\left( 1 + \sigma \Omega \right)}
         {\left(1 +  \Omega \right)^2}
\gg
  \frac
  {
  \left| 
    \tilde{B}(\vec{r}_\bot,\xi,\Omega ; A)
  \right|
  }
  { \left|
     \tilde{A}(\vec{r}_\bot,\xi,\Omega)
    \right|
  }
\label{condition-nonlinearity}
.
\end{eqnarray}

\end{description}

We use $\Omega$ instead of $\imath \partial_\tau$ for these conditions 
because conditions on the frequency components of the various terms are 
better defined than those for time derivatives.  

These conditions are the same as those required for the SEWA and SEEA
theories, with the SVEA conditions being a special case gained by setting
$\left| \Omega \right| \ll 1$ for the diffraction and nonlinearity conditions
-- implying that modulations in the envelope are so slow compared to the
carrier frequency that they can be neglected.  Note that backwardly
propagating behaviour has not been explicitly excluded, but since it would
appear as a modulation on the envelope $A$, it would be approximated away as
part of the evolution condition  (eqn. (\ref{condition-evolution})).

Note that the exact solution of eqn. (\ref{exact-BKP})
makes no reference to a particular choice of carrier phase $\psi_0$.  This
implies that once a solution for the propagation of a particular envelope has
been obtained, the problem has in fact been solved for a set of pulses 
(and initial conditions) based on
different carrier phases -- where that set is determined by the initial
envelope and some arbitrary choices of carrier phase $\psi_1 \in \left[ 0,
2\pi \right)$.  The final state is then given by the chosen $\psi_1$ combined
with the final form of the envelope.

\end{section}


\begin{section}{The $\chi^{(2)}$ nonlinear system}\label{chi2}

When modelling $\chi^{(2)}$ nonlinear systems we split the optical field 
into two or three parts, depending on whether a degenerate or non-degenerate
system is being treated.  For example, a parametric amplifier would have pump,
signal, and idler field components.  We then define an envelope $A_\alpha$ and
carrier $e^{\imath \Xi_\alpha}$, $\Xi_\alpha = \beta_{\alpha,0} z -
\omega_{\alpha,0} t + \psi_{\alpha,0}$ for each field component, and use a
separate propagation equation for each.  The total field is then the sum of
these different components:

\begin{eqnarray}
E &=& E_p + E_s + E_i
\ifthenelse{\boolean{BoolLong}}
{
\nonumber 
\\
&=& A_p e^{\imath \Xi_p} + A_p^* e^{-\imath \Xi_p} 
+ A_s e^{\imath \Xi_s} + A_s^* e^{-\imath \Xi_s} 
\nonumber 
\\
&&+ A_i e^{\imath \Xi_i} + A_i^* e^{-\imath \Xi_i} 
}
{
\nonumber 
\\
&=& A_p e^{\imath \Xi_p}
+ A_s e^{\imath \Xi_s} + A_i^* e^{-\imath \Xi_i} + \textrm{c.c.}
}
\label{eqn-Esum}
\end{eqnarray}

Because the wave equation eqn. (\ref{eqn-3DWE-time}) is linear in the electric
field, we can use eqn. (\ref{eqn-Esum}) in the theory of section \ref{theory},
and get a GFEA-like equation which is rather like the sum of three separate
copies of eqn. (\ref{exact-BKP}), but with a single polarization term.  In a
$\chi^{(2)}$ medium this polarization term is proportional to the square of
the field, and inspection of eqn. (\ref{eqn-Esum}) shows that this can be
expanded into a sum of many terms.  We can then split the multiple-field GFEA
equation into three separate GFEA-like equations, one for each envelope.  In
doing this we need to make sure to assign the pieces of the polarization
term appropriately to a suitable envelope equation, whilst discarding those
that are not resonant with a field carrier.

\begin{figure}
\includegraphics[width=85mm]{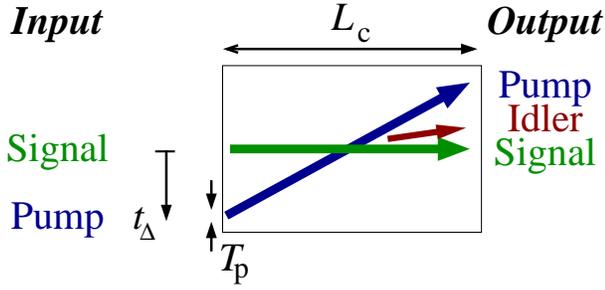}
\caption{ 
\label{Ftimes}
Pump timing offset (see section \ref{chi2}).   The pump pulse is 
injected into the crystal just before the signal pulse is reflected 
off the input mirror.  The faster moving signal pulse then catches
the pump pulse up about halfway through the crystal, and an idler pulse
is generated.
}
\end{figure}

Our chosen nonlinear crystal is congruent  LiNbO$_3$, for which we 
calculate refractive index and dispersion data from
the Sellmeier equations of Jundt \cite{Jundt-1997ol}.  
    We model the nonlinear polarization
    using the square of the total electric field, retaining the 
    parts resonant with our field carriers in the normal way.
Our OPO
simulations (see section \ref{opo}) assumed a pump frequency of
357.1 THz, with nominal signal and idler carrier frequencies of 257.5 THz and
99.6 THz respectively (wavelengths $0.84000 \mu$m, $1.16500 \mu$m, $3.0110
\mu$m).  This means the pump pulse will travel through LiNbO$_3$ more slowly
than the signal and it therefore needs to be injected into the
crystal ahead of it (see fig. \ref{Ftimes}).  When the two overlap,
an idler is generated by the nonlinear interaction, and the three pulses then
continue to interact with each other as they propagate through the crystal.
Note that our
ideal non-degenerate parametric amplifier simulations (see section
\ref{idealchi2}) use the same field frequencies, but idealise the crystal
parameters by setting the group velocities and dispersions to zero.



\begin{subsection}{System Scalings}\label{scaling}

In a typical experiment, the crystal length would be fixed, as would any
properties defined by its design, such as periodic poling.  If we were to
investigate this case for a range of pulse durations, there would naturally be
differences between the results, even within the SVEA.  For example, the
relative pulse broadening caused by travelling through a $1000 \mu$m crystal
is greater for a 12fs pump pulse than for a 48fs one.  Similarly, a
fixed timing offset for injection would have different
effects; and a fixed pump pulse power would generate different strengths of
nonlinear interaction for different pulse lengths.  All these effects would
confuse any attempt at a systematic comparison of the few cycle effects in the
models we consider.

Therefore, in order to isolate specific few-cycle effects, we must scale the
pump pulse FWHM $\mathcal{T}_\textrm{p}$, crystal length $L_\textrm{c}$, pump
pulse energy $\mathcal{W}$, and pump timing offset $t_\Delta$ in such a way as
to ensure the effects of group velocity, pump timing, and nonlinearity occur
in the same proportions to one another over the range of pulse lengths.  

We can work out an appropriate scaling  by examining a
simple version of the propagation equation (eqn.\ref{exact-BKP}), where we
write the group velocity prefactors as  $B_1$, the second order dispersion
prefactors as $B_2$, and the polarization terms as $C A^2$.  To assist us with
the scaling process we also write $\xi=r^{-f}\xi'$, $\tau=r^{-g}\tau'$, and
$A=r^hA'$, where $r$ is the scaling factor. Our simple propagation equation is
therefore

\begin{eqnarray}
 r^{h+f} \partial_{\xi'} A' 
 &=&
 r^{h+g} B_1 \partial_{\tau'} A'
+
 r^{h+2g} B_2 \partial_{\tau'}^2 A'
+
 r^{2h} C A'^2
.
\label{e-simpleprop}
\end{eqnarray}

We can easily match the LHS term with the polarization term by setting
$f=h$; but then we must choose either $h=g$ to match group velocities, or
$f=2g$ to match the second order dispersion -- we cannot match both.  For our
chosen OPO situation (see section \ref{opo}), it is best to match the group
velocity terms, which control how long the pump and signal pulses overlap --
in general, the dispersion has a much smaller effect.

We take our reference situation to be 
a $20$nJ $24$fs FWHM pump pulse propagating through a $500\mu$m crystal,
with a pump timing offset of $48$fs.  For the chosen parameter scaling 
($f=g=h$)

\begin{eqnarray}
  \frac{\mathcal{T}_\textrm{p}}
       {24\textrm{fs}}
 = 
  \frac{L_\textrm{c}}
       {500\mu\textrm{m}}
 =
  \frac{t_\Delta}
       {48\textrm{fs}}
 =
  \frac{20\textrm{nJ}}
       {\mathcal{W}}
.
\label{scalingsummary}
\end{eqnarray}

This removes the gross effects caused by reducing the pulse
duration, so that if there are no few cycle effects, and second order 
dispersion is negligible, {\em each pulse length
should give identical results}.

First, the crystal length $L_\textrm{c}$ is scaled in direct proportion to the
pulse length $\mathcal{T}_\textrm{p}$.  This means that the relative group
velocities are scaled so that the pump and signal pulses spend the same 
relative time overlapping, leading to an equivalent nonlinear effect.

Second, and related to the previous point, the pump timing offset $t_\Delta$
is reduced in direct proportion to the crystal length (and hence pulse
length).  This means the pump-signal pulse cross-over occurs at the same
relative point, so the effect of the pump-signal overlap is similar.  If a
48fs pump pulse in a $1000 \mu$m crystal arrives $96$ns before the signal, the
crosover occurs at about the $500\mu$m half-way mark -- just as for a 12fs
pulse in a $250 \mu$m crystal, where a $t_\Delta=24fs$ gives a crossover at
the $125 \mu$m half-way mark.

Third, a shorter crystal reduces the effect of the nonlinearity,  so the pump
pulse energy is scaled up as the inverse of the reduction in crystal length --
a half length crystal needs pulse energies $\mathcal{W}$ twice as large -- a
10nJ 48fs pulse in its $1000 \mu$m crystal with see the same nonlinear effect
as a 20nJ 24fs pulse in its $500 \mu$m crystal.  Note that the pulse energy
depends on the area of the pulse, and so depends on its length, so that 
$\mathcal{W} \propto \left|A\right|$.

We could choose to make the scaling perfect, by also scaling the crystal
parameters.  
    If we scale the crystal dispersion with $B_2=r^{-g}B_2'$, the relative 
    amount of pulse spreading changes to become 
    the same for each 
simulation -- e.g.  if the 48fs pulse
widens by 10\% in a $1000 \mu$m crystal, the 12fs will also widen by 10\% in
its $250 \mu$m crystal.  We did a set of SVEA simulations on this basis, and
as expected saw identical pulse profiles regardless of the chosen pulse
length.  However, we chose not to use this perfect scheme for the bulk of our
OPO simulations because it is far from being experimentally practical.

\end{subsection}

\end{section}


\begin{section}{Ideal Parametric Interactions}\label{idealchi2}

A parametric amplifier is a single-pass device: pump and signal pulses are
injected into one end of the nonlinear crystal, they interact within it,
then exit at the far end.  However, because real nonlinear
crystals (such as LiNbO$_3$) tend to have significant dispersion, very short
pulses quickly spread out, making them difficult to create, and reducing the
few-cycle effects we aim to study.

In order to demonstrate clearly the nature of few-cycle effects in
$\chi^{(2)}$ materials, in this section we investigate an ideal case by 
setting the dispersion to zero, and make the group velocity the same at all
frequencies.  This means that $\sigma=1$, so the ``few-cycle''
polarization prefactor $\mathcal{K}$ is identical for both the SEWA and GFEA
theories. Note that it is difficult to do no-dispersion simulations over long
times because pulse self-steepening causes both the numerical integration and
the theoretical approximations to break down.

We inject Gaussian pump and signal pulses at exactly the same time (i.e. 
$t_\Delta=0$), with the same width.  They then travel down the crystal with
maximum overlap, interacting all the way.  Other parameters are fixed by the
scaling rules from section \ref{scaling}.  Further, when {\em graphing}
results for the figures, we scale the times for each pulse length to the 6fs
case (e.g. for a 24fs pulse, ``$\tau=10$'' corresponds to $40$fs), and scale
the pulse intensities in proportion to their initial intensities.  This means
that graphs of the initial conditions for a range of pulse lengths
would be identical.

Finally, note that
in these ideal results, the nonlinear interaction is ``strong'', with
significant transfer of energy between the fields.

%
%
%
%
%

\ifthenelse{\boolean{BoolSHG}}{

\begin{subsection}{Second Harmonic Generation (SHG)}\label{idealchi2-shg}

\begin{figure}
\includegraphics[width=85mm]{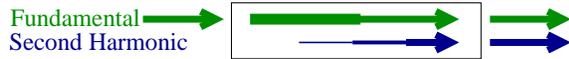}
\caption{ 
\label{Fshg-diagram}
SHG: 
Second Harmonic Generation (section \ref{idealchi2-shg}).  The thickness
of the arrows is intended to give an indication of how the energy
of the field components changes during propagation through the crystal.
}
\end{figure}

\ifthenelse{\boolean{BoolLong}}
{
\begin{figure}
\includegraphics[width=53mm,angle=-90]{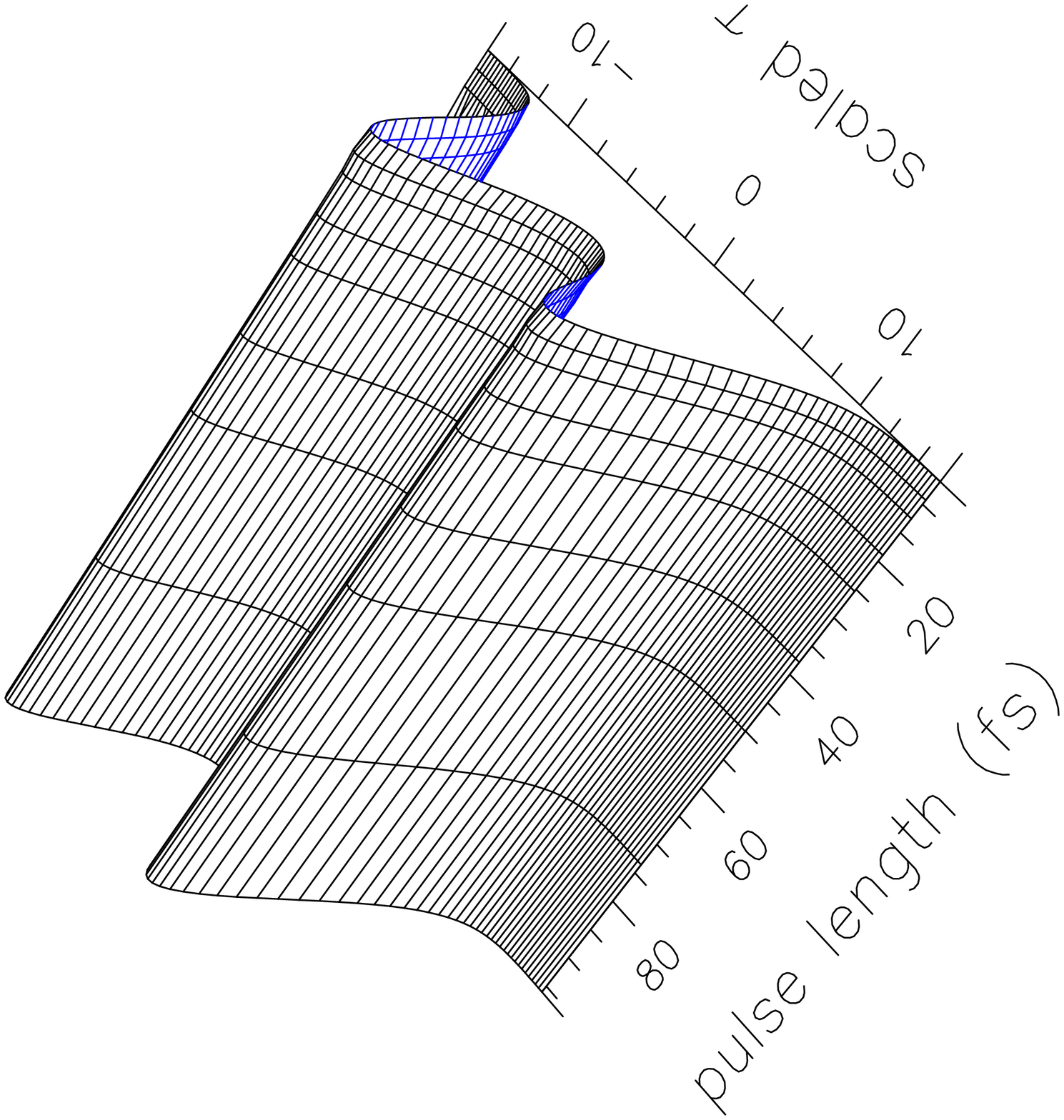}
\includegraphics[width=53mm,angle=-90]{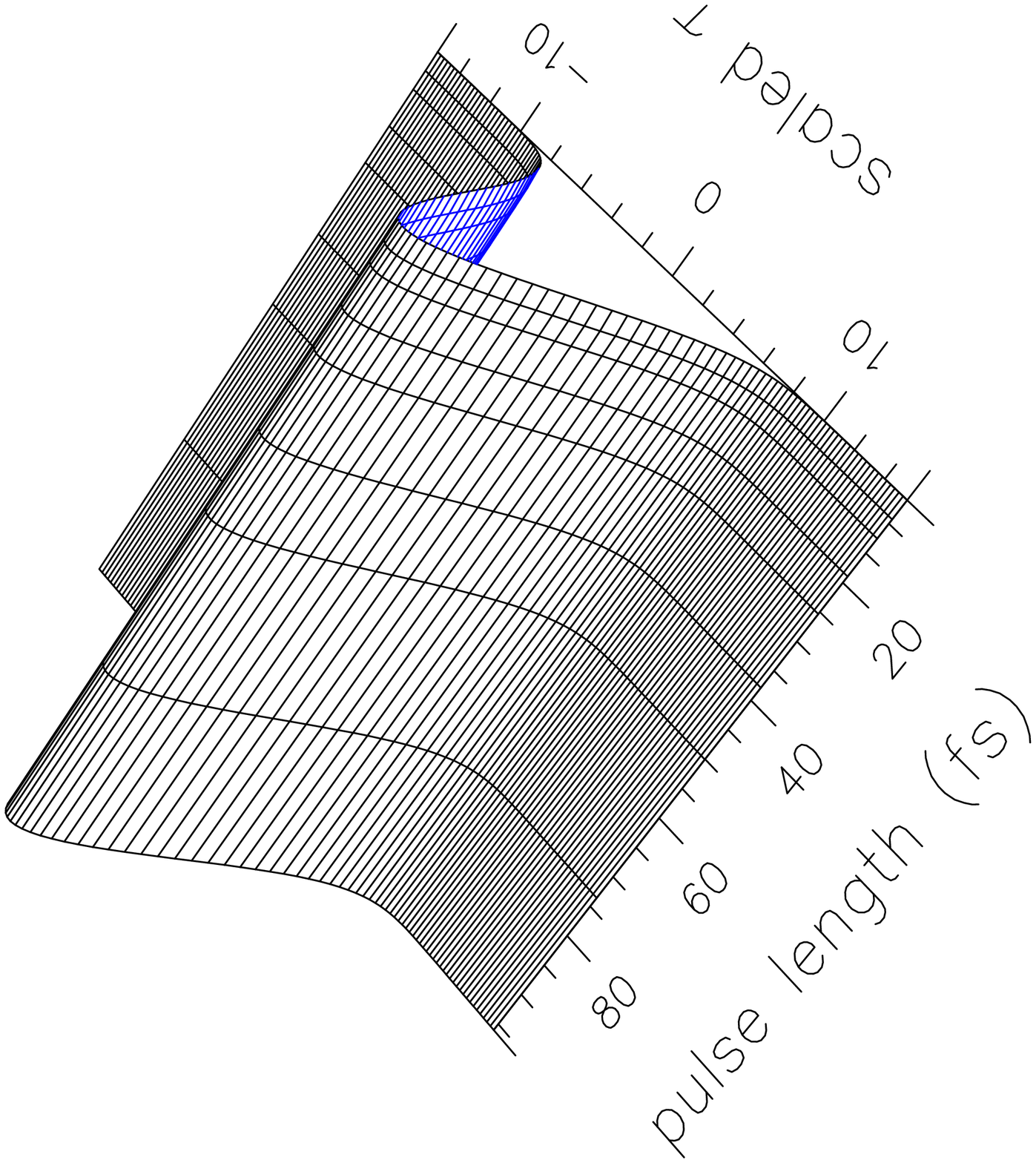}
\caption{ 
\label{Fshg-single-SVEA}
SHG:
Scaled pulse envelopes $\left|A\right|^2$ from the SVEA,
on exit from the ideal dispersionless crystal.
Top is fundamental, bottom is second harmonic.
}
\end{figure}
}

\begin{figure}
\includegraphics[width=53mm,angle=-90]{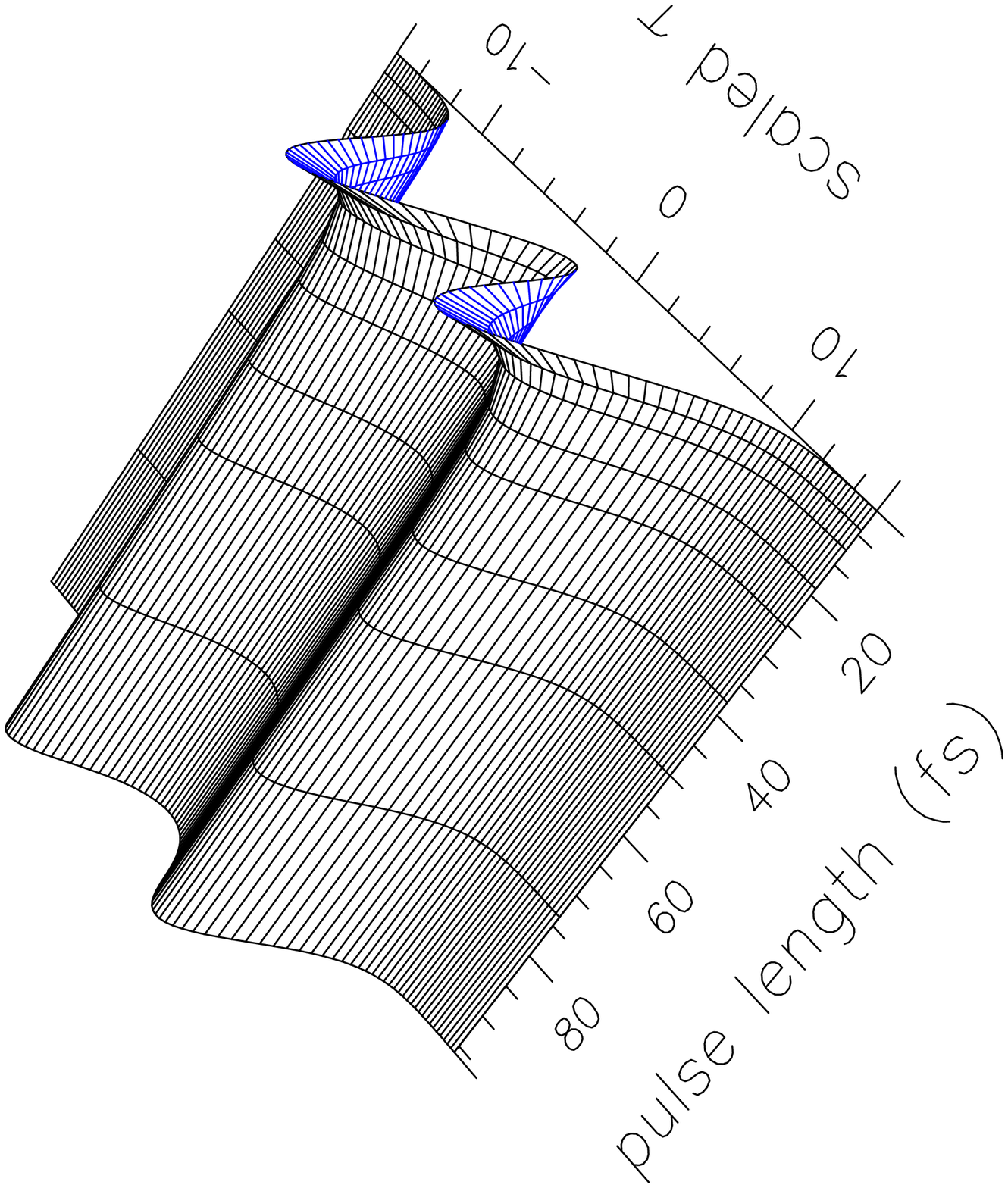}
\caption{ 
\label{Fshg-single-fundamental}
SHG:
Scaled fundamental pulse envelopes $\left|A\right|^2$ from the GFEA,
on exit from the ideal dispersionless crystal.
The SVEA results for {\em all} pulse lengths are essentially identical to 
the 96fs result.  
}
\end{figure}

\begin{figure}
\includegraphics[width=53mm,angle=-90]{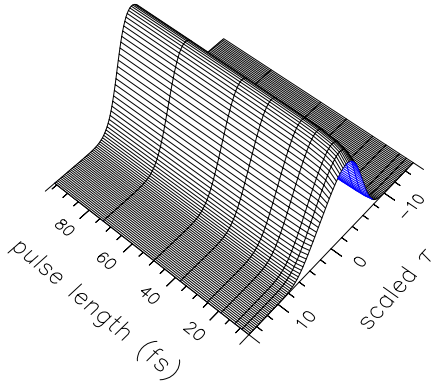}
\caption{ 
\label{Fshg-single-2harmonic}
SHG:
Scaled second harmonic pulse envelopes from the GFEA, 
see \protect{fig. \ref{Fshg-single-fundamental}}.
The SVEA results for {\em all} pulse lengths are essentially identical to 
the 96fs result.  
}
\end{figure}

In second harmonic generation the nonlinearity causes a field at one
frequency (the fundamental) to generate one at twice that frequency (the
second harmonic).  In our 48fs simulations, a 10nJ pulse is injected in at
the fundamental frequency (see fig.  \ref{Fshg-diagram}).

In Figs. \ref{Fshg-single-fundamental}, \ref{Fshg-single-2harmonic} we can see
comparisons of the pulses resulting from a single pass through the crystal. 
Although the plotted (and scaled) $\left|A\right|^2$ of the pulses are rather
similar; the GFEA theory gives noticeably different results for the shortest
pulses, unlike the SVEA theory.  Note that the vertical scales on the two figures (not shown) are rather different.

However, this apparent similarity in $\left|A\right|^2$ is a little misleading
because the underlying phase profiles of the envelopes are very different, as
can be seen in (see fig. \ref{Fshg-single-phase}).  More features, primarily
extra oscillations,  do develop at shorter pulse durations, but generally the
comparisons remain similar to those shown.

\begin{figure}
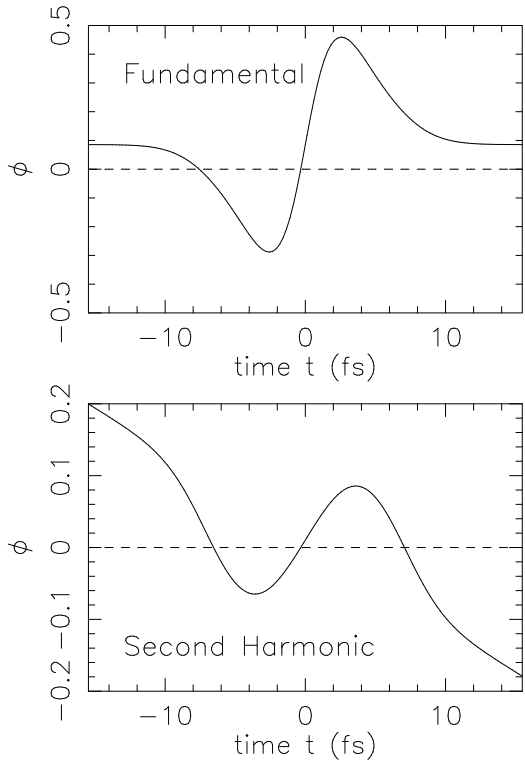

\includegraphics[width=50mm,angle=-90]{shg-Single_d1-signal-0125u-0000pp-pc}
\includegraphics[width=50mm,angle=-90]{shg-Single_d1-pump-0125u-0000pp-pc}
\caption{ 
\label{Fshg-single-phase}
SHG:
envelope-phase profiles on a scaled $\tau$, 
for an 18fs pulse duration. 
SVEA  ({--~--~--}),
GFEA  ({------}).
NB: the label ``time t'' should read ``scaled $\tau$''.
}
\end{figure}

\end{subsection}

} 

%
%
%
%
%

\ifthenelse{\boolean{BoolDPA}}{

\begin{subsection}{Degenerate Parametric Amplification (DPA)}\label{idealchi2-dpa}

\begin{figure}
\includegraphics[width=85mm]{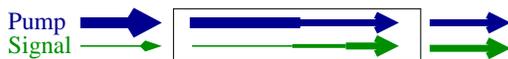}
\caption{ 
\label{Fdpa-diagram}
DPA: 
 Degenerate Parametric Amplification (section \ref{idealchi2-dpa}).  
The thickness
of the arrows is intended to give an indication of how the energy
of the field components changes during propagation through the crystal.
}
\end{figure}

A degenerate parametric amplifier uses energy from a pump pulse to amplify
that of a signal pulse at half the frequency ($\omega_p = 2 \omega_s$, see
fig.  \ref{Fdpa-diagram}).  For our 24fs reference simulations, a 20nJ pulse
is injected at the pump frequency, and a signal pulse of 10pJ is amplified;
the nonlinear interaction is very strong.

In Figs. \ref{Fdpo-single-signal}, \ref{Fdpo-single-pump} we can see
comparisons of the pulses resulting from a single pass through the crystal. 
Note that the intensity profiles of the pulses are very similar, but the GFEA
theory gives different results to the SVEA for the shortest pulses.  

\ifthenelse{\boolean{BoolLong}}
{
\begin{figure}
\includegraphics[width=53mm,angle=-90]{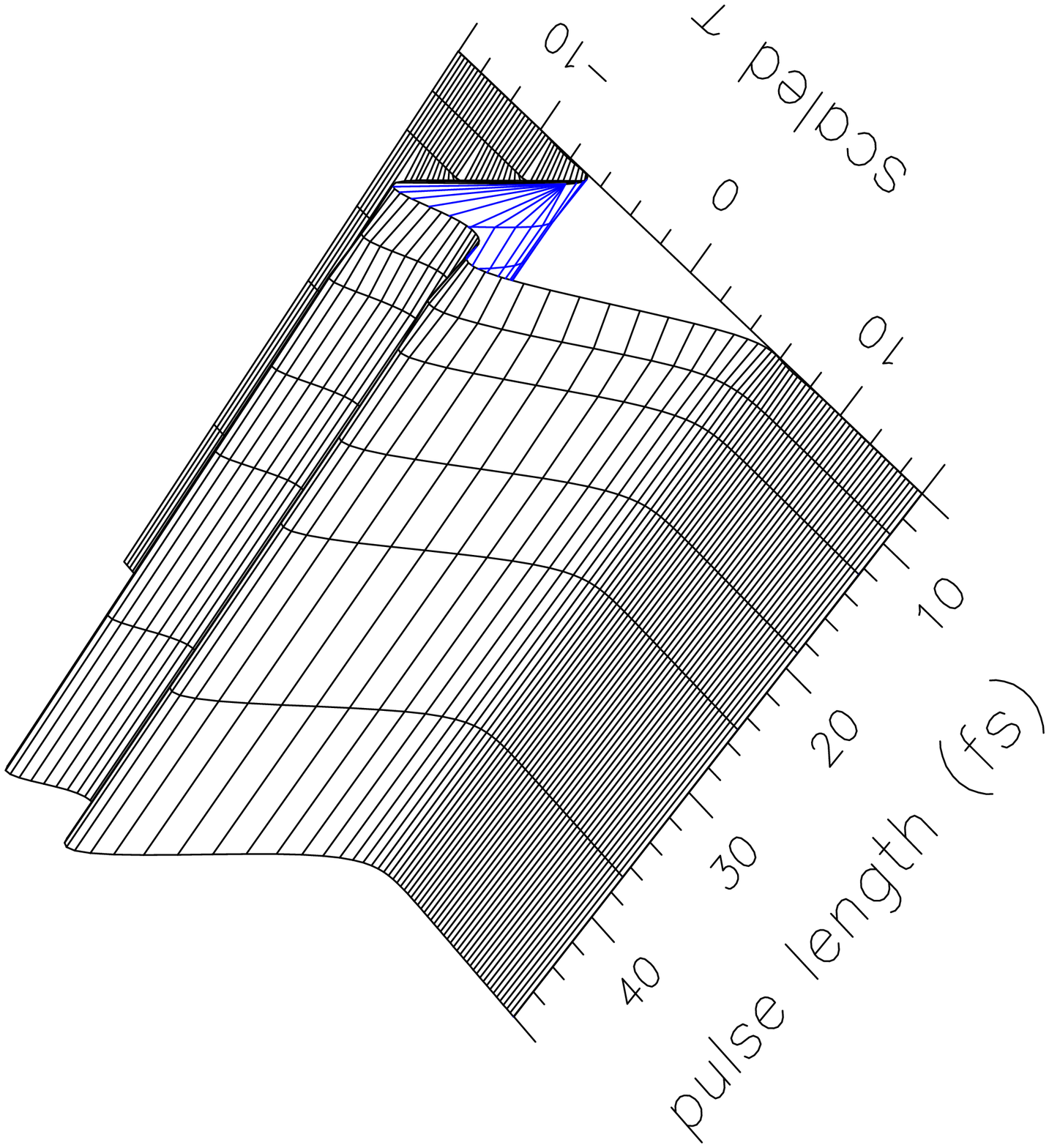}
\includegraphics[width=53mm,angle=-90]{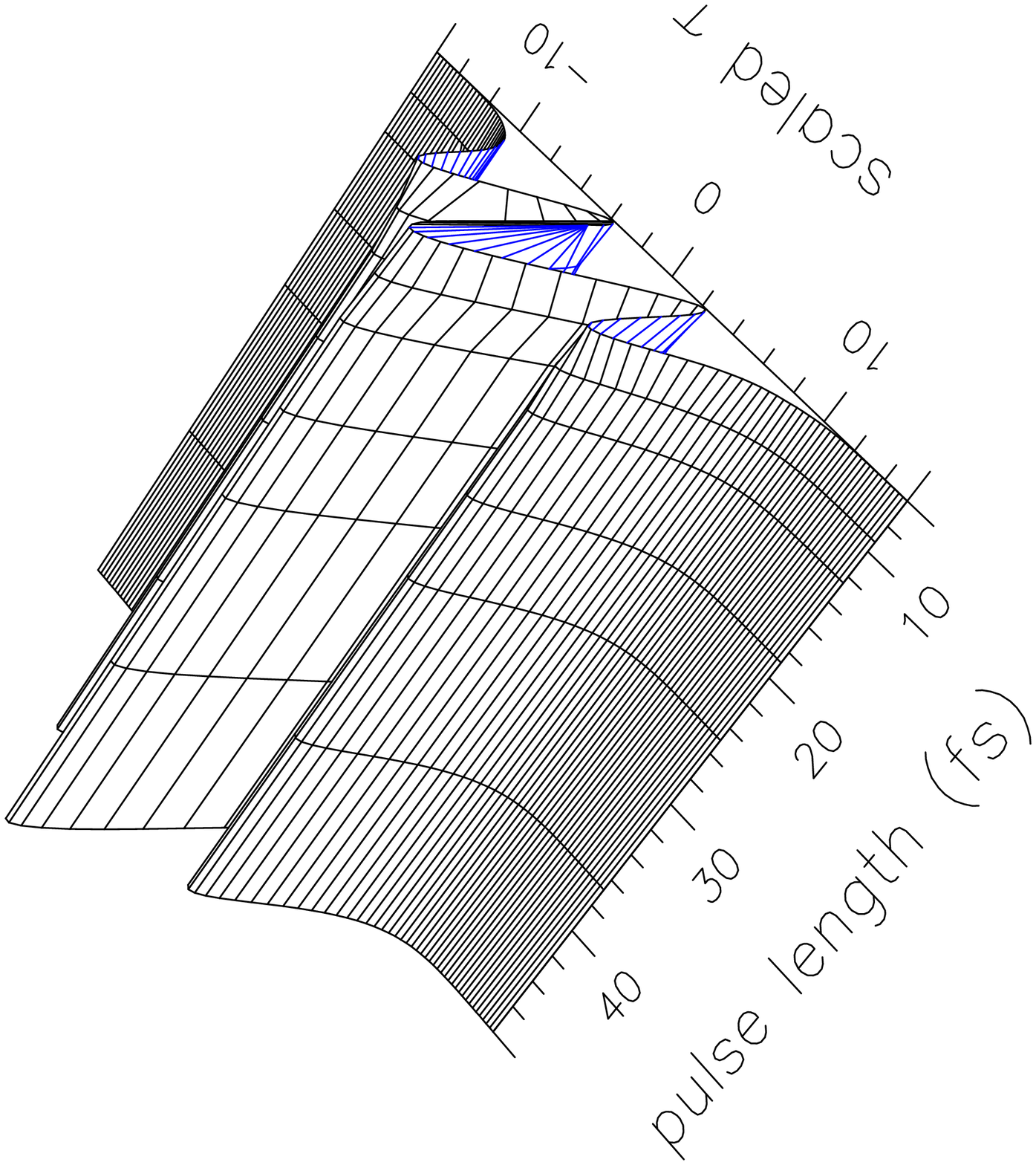}
\caption{ 
\label{Fdpo-single-SVEA}
DPA:
Scaled pulse envelopes $\left|A\right|^2$ from the SVEA,
on exit from the ideal dispersionless crystal. 
Top is signal, bottom is pump. 
}
\end{figure}
}

\begin{figure}
\includegraphics[width=53mm,angle=-90]{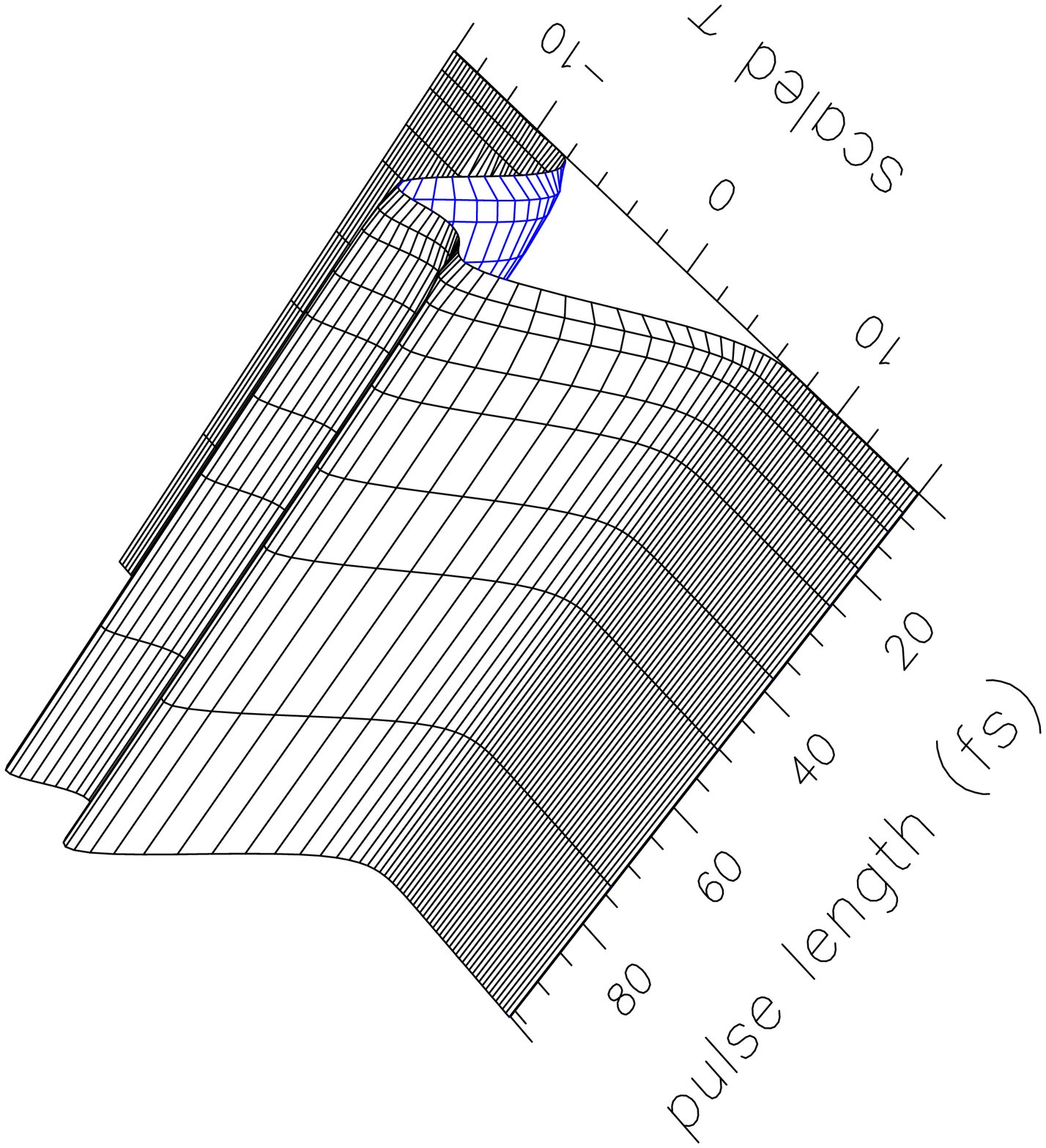}
\caption{ 
\label{Fdpo-single-signal}
DPA:
Scaled signal pulse envelopes $\left|A\right|^2$ from the GFEA, 
on exit from the ideal dispersionless crystal. 
The SVEA results for {\em all} pulse lengths are essentially identical to 
the 96fs result.  
}
\end{figure}

\begin{figure}
\includegraphics[width=53mm,angle=-90]{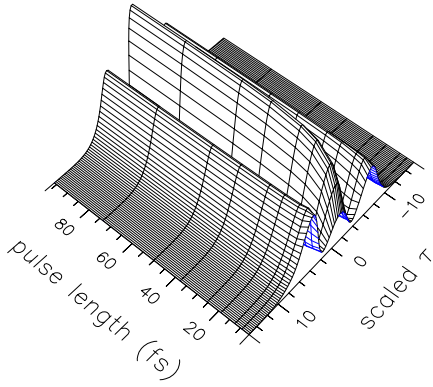}
\caption{ 
\label{Fdpo-single-pump}
DPA:
Scaled pump pulse envelopes from the GFEA, 
see \protect{fig. \ref{Fdpo-single-signal}}.
The SVEA results for {\em all} pulse lengths are essentially identical to 
the 96fs result.  
}
\end{figure}

However, this apparent similarity
in $\left|A\right|^2$ is a little misleading because the underlying phase
profiles of the envelopes are very different, as can be seen in 
fig. \ref{Fdpo-single-phase}.  The change in phase of the pump pulse in both
the SVEA and GFEA case is due to the presence of nodes that develop to either
side of its centre.  More features, primarily extra oscillations,  do develop
at shorter pulse durations, but generally the comparisons remain similar.  

\begin{figure}
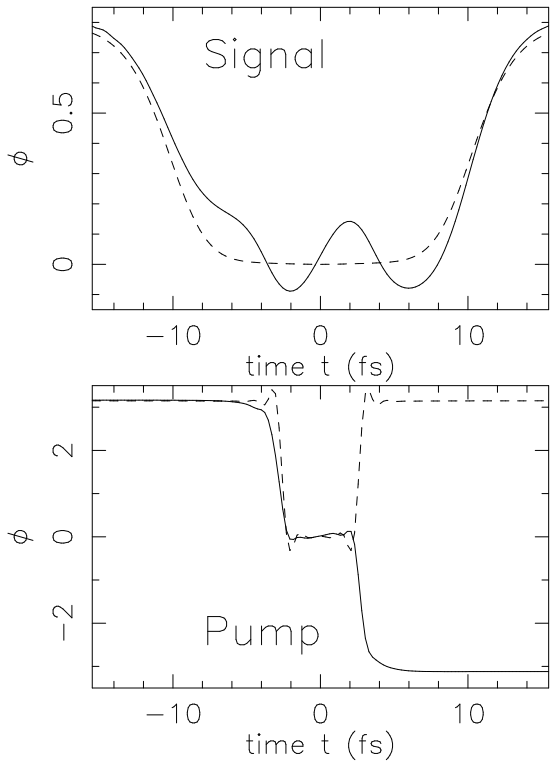

\includegraphics[width=50mm,angle=-90]{dpo-Single_d1-signal-0125u-0000pp-pc}
\includegraphics[width=50mm,angle=-90]{dpo-Single_d1-pump-0125u-0000pp-pc}
\caption{ 
\label{Fdpo-single-phase}
DPA:
envelope-phase profiles on a scaled $\tau$, 
for an 18fs pulse duration. 
SVEA  ({--~--~--}),
GFEA  ({------}).
.
}
\end{figure}

\end{subsection}

} 

%
%
%
%
%

\ifthenelse{\boolean{BoolDPD}}{

\begin{subsection}{Degenerate Parametric De-amplification (DPD)}\label{idealchi2-dpd}

\begin{figure}
\includegraphics[width=85mm]{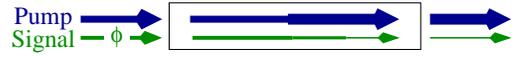}
\caption{ 
\label{Fdpd-diagram}
DPD: 
Degenerate Parametric De-amplification (section \ref{idealchi2-dpd}). The thickness
of the arrows is intended to give an indication of how the energy
of the field components changes during propagation through the crystal.
}
\end{figure}

This is the same situation as for degenerate parametric amplification, but we
have changed the relative phase between the pump pulse and signal so that the
input signal pulse is de-amplified (see fig. \ref{Fdpd-diagram}).  We are
interested in this case because the de-amplification is very sensitive to the
phase relationship between the pulses, and the few-cycles terms modify the
phases of the pulses.  Our SVEA model of the situation behaves in the
expected way, with the signal pulse decaying away to zero as it propagates
through the crystal and transfers its energy to the pump.

\ifthenelse{\boolean{BoolLong}}
{
\begin{figure}
\includegraphics[width=50mm,angle=-90]{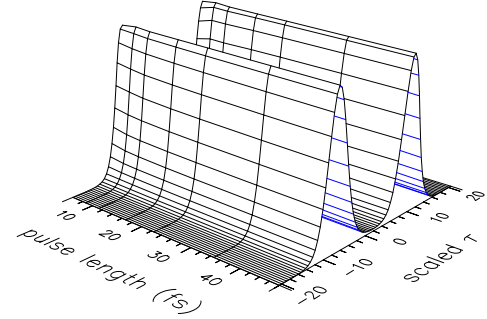}
\caption{ 
\label{Fdpd-single-SVEA}
DPD:
Output SVEA signal intensities for $\phi_s=\pi/4$ and $\phi_p=0$ for a range
of pulse durations: 
peak value $\left| A \right|^2 \approx 4 \times 10^2$;
}
\end{figure}
}

\begin{figure}
\includegraphics[width=50mm,angle=-90]{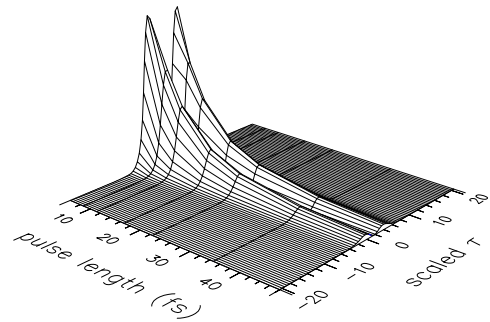}
\caption{ 
\label{Fdpo-single-phase-cf}
DPD:
Output GFEA signal intensities for $\phi_s=\pi/4$ and $\phi_p=0$ for a range
of pulse durations, peak value $\approx 4 \times 10^5$.
The intensities and times are scaled in our usual way. 
The SVEA results are nearly identical for all pulse lengths, and 
would not show up on the scale of this GFEA graph: their output
signal intensity consists of two peaks (at $\tau \approx \pm 10$), 
with a height less than 1/30th of that of the smallest (48fs) GFEA peaks.
}
\end{figure}

However the new terms in the GFEA propagation equation adjust the phases of
both the pump and signal fields as they propagate, moving them away from an
exact match to the de-amplification criteria. As the mismatch increases, the
de-amplification process is reversed, and instead the usual amplification
takes over.  Fig. \ref{Fdpo-single-phase-cf} shows the output signal pulse
profiles on exit from the crystal, note the enormous difference in the
few-cycle dependence of the SVEA and GFEA theories -- both in character and
amplitude.  The SVEA profile is just the residual input signal pulse which has
not yet been fully ``de-amplified''; this residue can also be seen as a
shoulder on the wings of the GFEA profiles if plotted on a $\log$ scale.  The
large and very visible double peak in the GFEA graph is a result of the finite
pulse length on the nonlinear polarizations term -- as can be seen in
eqn. (\ref{exact-BKP-altB}), this alters the phase profile of the nonlinear
effect, and hence the pulses, which means the signal no longer satisfies the
exact criteria for de-amplification.  For our chosen input pulse powers, this
effect persists for initial phase mismatches up to $\sim 0.01$ radians, after
which the amplification of the signal pulse has largely swamped the anisotropy
induced by the few-cycle effects.  Of course the effect is most visible when
the interaction has been strong enough for the input component of the signal
pulse to (almost) completely disappear.

\ifthenelse{\boolean{BoolLong}}
{
\begin{figure}
\includegraphics[width=50mm,angle=-90]{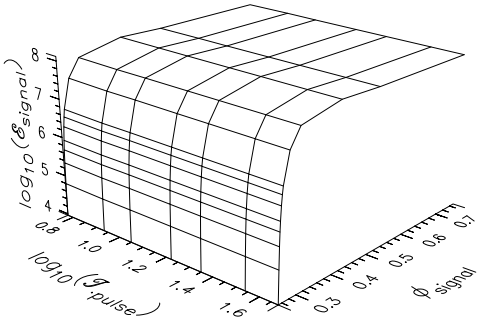}
\caption{ 
\label{Fdpo-single-phase-ecf-S}
DPD:
Output SVEA pulse energies ($\mathcal{E}_{\mathrm{signal}} =
  \int{\left| A_{\mathrm{signal}} \right|^2 d\tau}$, arbitrary units) 
for a range of 
initial signal (envelope) phases $\phi_s$ 
and pulse lengths $\mathcal{T}_{\mathrm{p}}$.
}
\end{figure}
}

\begin{figure}
\includegraphics[width=50mm,angle=-90]{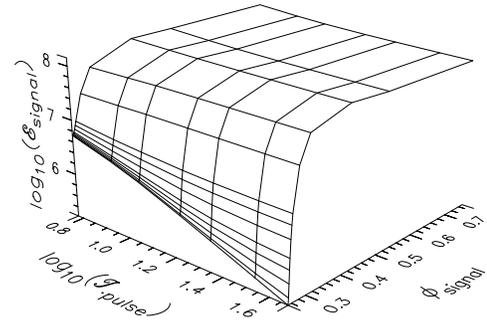}
\caption{ 
\label{Fdpo-single-phase-ecf}
DPD:
Output GFEA pulse energies ($\mathcal{E}_{\mathrm{signal}} =
  \int{\left| A_{\mathrm{signal}} \right|^2 d\tau}$, arbitrary units) 
for a range of 
initial signal (envelope) phases $\phi_s$ 
and pulse lengths $\mathcal{T}_{\mathrm{p}}$.
The intensities and times are scaled in our usual way. 
The SVEA results are nearly identical for all pulse lengths, and 
differ from the 48fs (i.e. $\log_{10}(48)=1.68$) results in 
that the near $\phi_{\mathrm{signal}}=\pi/4$ give significantly
lower energies -- down to $10^4$ rather than $10^5$ for $\pi/4$.
The graphed signal phases in multiples of $\pi$ are are 0.250, 0.251, 
0.252, 0.253, 0.254, 0.256, 
0.258, 0.260, 0.275, 0.300, 0.400, ..., 2.000
}
\end{figure}


In fig. \ref{Fdpo-single-phase-ecf} we see how the behaviour changes both with
pulse length and initial phase match -- we see perfect de-amplification in the
SVEA model, but the steeply sloping $\phi_{\mathrm{signal}}=\pi/4$ line
clearly demonstrates the effects of a finite pulse length.  Of course as the
$\phi_{\mathrm{signal}}$ moves further from $\pi/4$, the SVEA model no longer
undergoes perfect de-amplification, and so the difference between
the two models becomes small.

\end{subsection}

} 

%
%
%
%
%

\ifthenelse{\boolean{BoolNPA}}{

\begin{subsection}{Non-degenerate Parametric Amplification (NPA)}\label{idealchi2-npa}

\begin{figure}
\includegraphics[width=85mm]{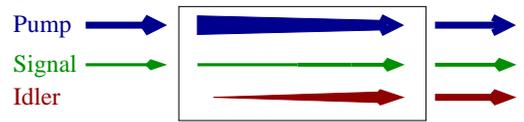}
\caption{ 
\label{Fnpa-diagram}
NPA: 
Non-degenerate Parametric Amplification (section \ref{idealchi2-npa}). 
The thickness
of the arrows is intended to give an indication of how the energy
of the field components changes during propagation through the crystal.
}
\end{figure}

We consider first a non-degenerate parametric amplifier with pump, signal, and
idler frequencies such that $\omega_p \rightarrow \omega_s + \omega_i$, and
$\omega_s \neq \omega_i$.  In the 24fs reference case, the initial pump energy
is 20nJ and the initial signal energy is 10pJ, with a negligible (but
finite) idler.  For other pulse durations, the energies were scaled according
to eqn. (\ref{scalingsummary}).  Fig. \ref{Fnpa-diagram} shows how, according
to the GFEA, the idler pulse intensity profiles $\left|A_i\right|^2$ generated
in a single pass of the crystal vary with pulse duration.  The profiles show
little variation with pulse duration except for the shortest pulses 
($\tau \lesssim 20$), where 
distortion is evident; the signal and pump profiles show deviations of a
comparable magnitude.

\ifthenelse{\boolean{BoolLong}}
{
\begin{figure}
\includegraphics[width=53mm,angle=-90]{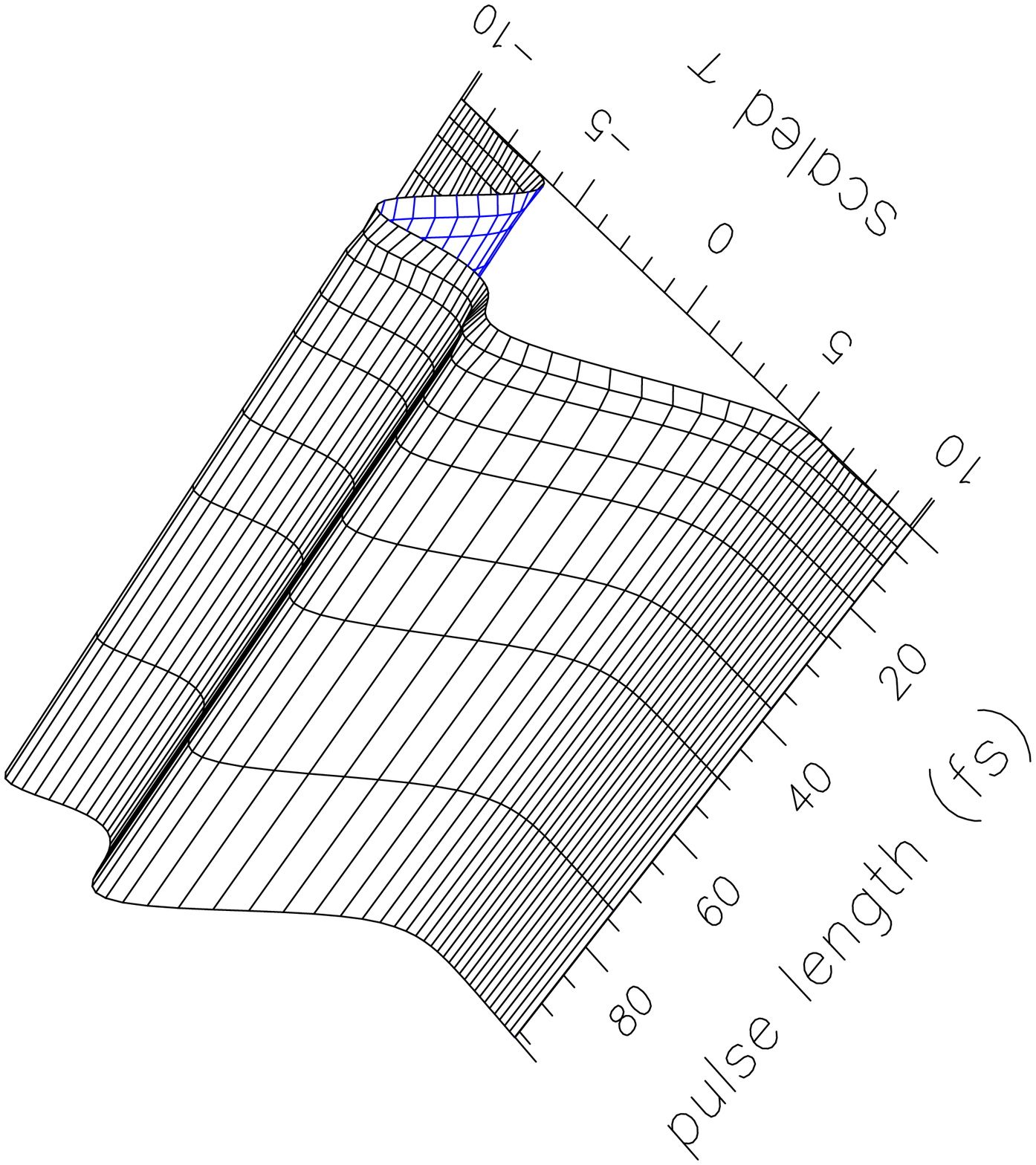}
\includegraphics[width=53mm,angle=-90]{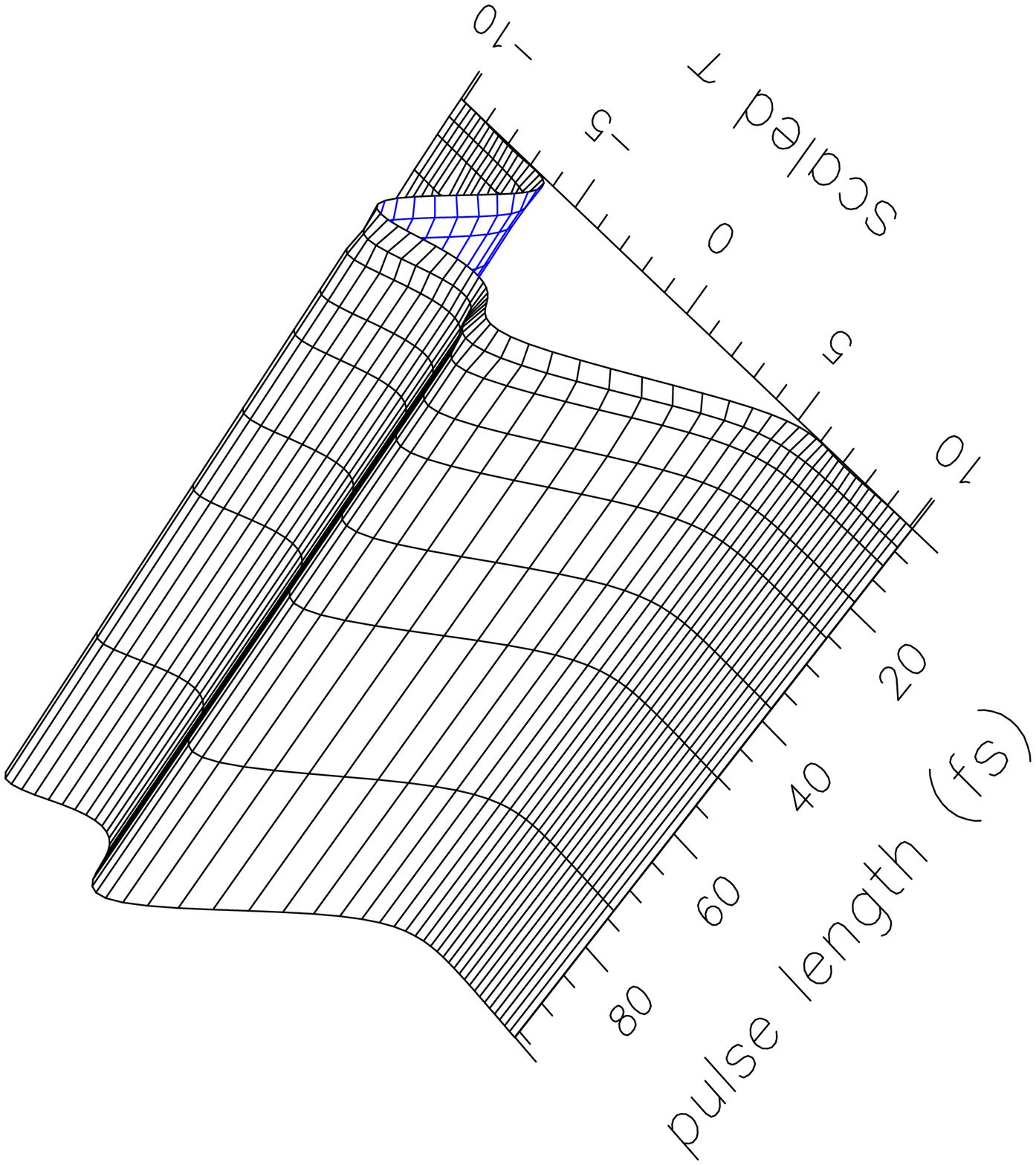}
\includegraphics[width=53mm,angle=-90]{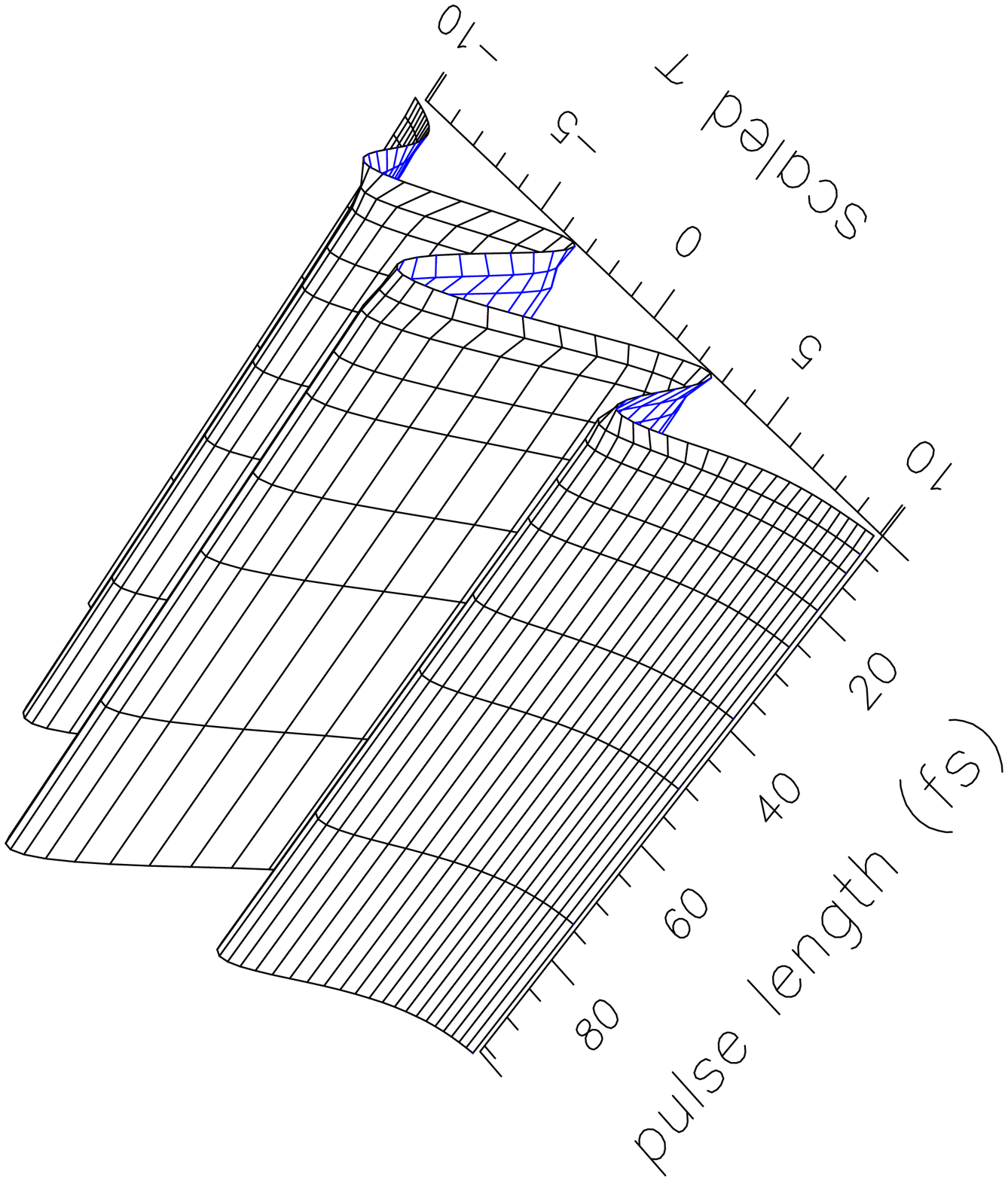}
\caption{ 
\label{Fnpo-single-SVEA}
NPA: 
Scaled SVEA pulse envelopes $\left|A\right|^2$ on exit from the ideal 
dispersionless crystal. 
Top to bottom is idler, signal, pump.
}
\end{figure}
}

\begin{figure}
\includegraphics[width=53mm,angle=-90]{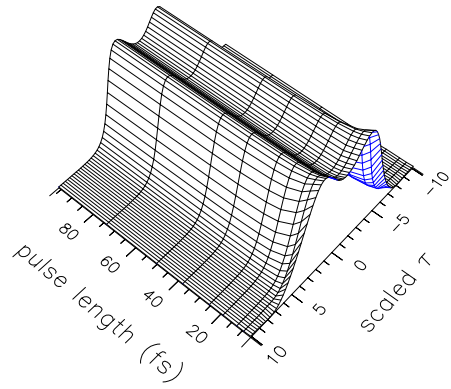}
\caption{ 
\label{Fnpo-single-idler}
NPA: 
Scaled GFEA idler pulse envelopes $\left|A\right|^2$ on exit from the ideal 
dispersionless crystal. 
The SVEA results for {\em all} pulse lengths are essentially identical to 
the 96fs result.  
}
\end{figure}

\ifthenelse{\boolean{BoolLong}}
{
In fig. \ref{Fnpo-single-idler}, \ref{Fnpo-single-signal},
\ref{Fnpo-single-pump} we can see equivalent comparisons for the signal and 
pump intensity profiles.
}
{}

\ifthenelse{\boolean{BoolLong}}
{
\begin{figure}
\includegraphics[width=53mm,angle=-90]{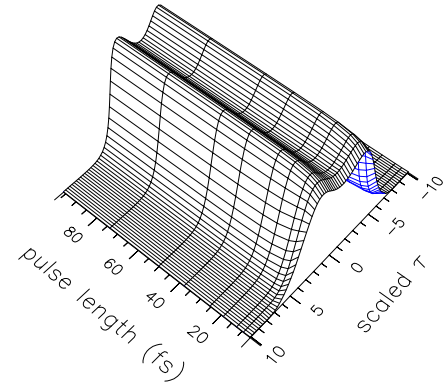}
\caption{ 
\label{Fnpo-single-signal}
NPA: 
Scaled GFEA signal pulse envelopes, see 
\protect{fig. \ref{Fnpo-single-idler}}.  
The SVEA results for {\em all} pulse lengths are essentially identical to 
the 96fs result.  
}
\end{figure}

\begin{figure}
\includegraphics[width=53mm,angle=-90]{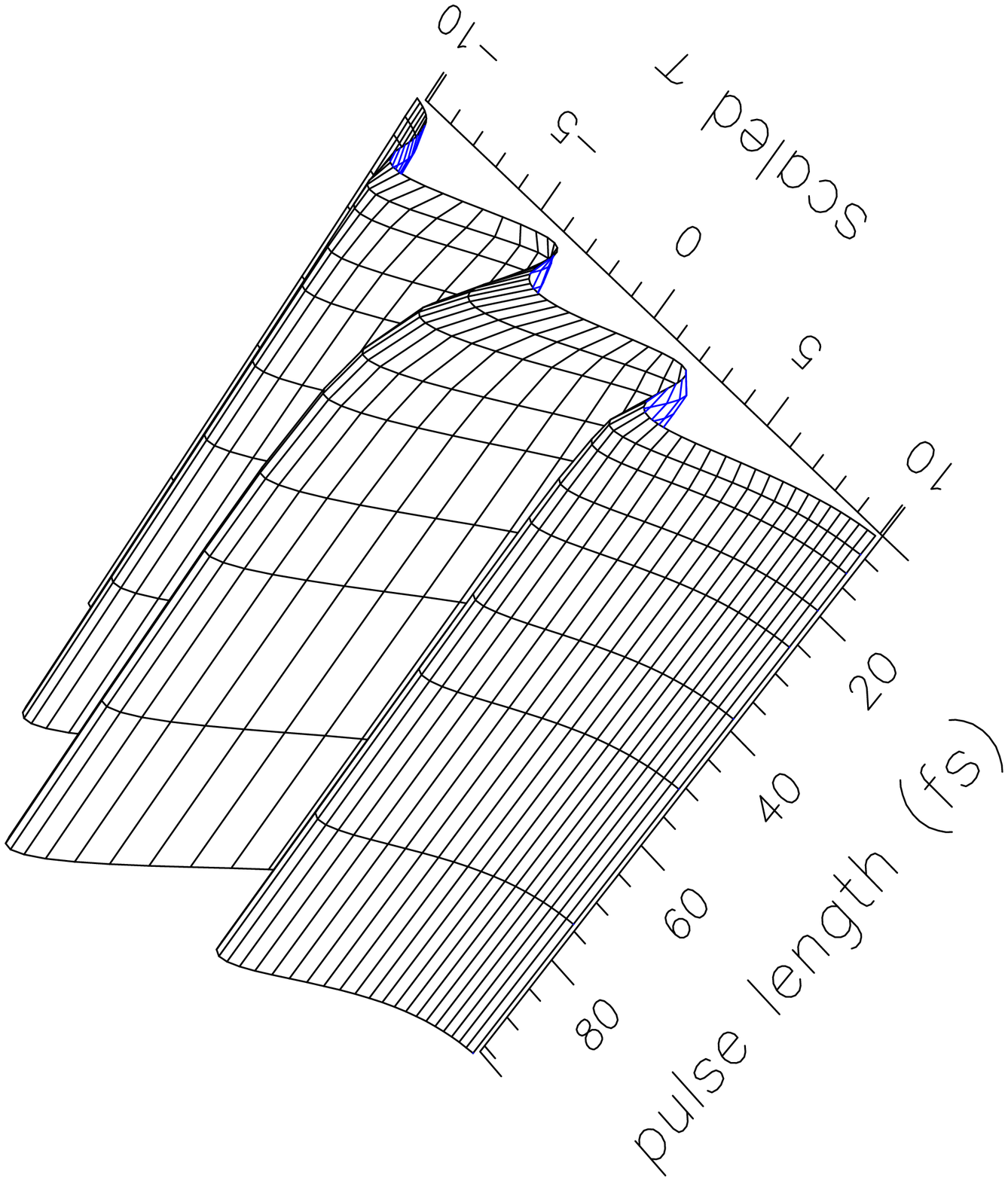}
\caption{ 
\label{Fnpo-single-pump}
NPA:
Scaled GFEA pump pulse envelopes, see \protect{fig. \ref{Fnpo-single-idler}}. 
The SVEA results for {\em all} pulse lengths are essentially identical to 
the 96fs result.  
}
\end{figure}
}

More dramatic effects appear in the phase profiles: in fig. 
\ref{Fnpo-single-phase}, the phases of the pulse envelopes at pulse durations
of 18fs and 96fs are shown with the phase distortions due to the finite pulse
lengths (see eqn. (\ref{exact-BKP-altB})).  As the pulse duration shortens, the
principal effect is to increase the magnitude of the phase distortion,  leaving
the shape of each profile largely unchanged; however more complex phase
oscillations develop for the shortest pulses.  At 96fs, the profiles show a
smaller distortion, and are tending towards the long-pulse SVEA limit.  In
this limit, the profiles are essentially flat, although the pump field
develops nodes which give rise to a step-like change in the phase.

\begin{figure}
\includegraphics[width=31mm,angle=-90]{npo-Single_d1-idler-pc}
\includegraphics[width=31mm,angle=-90]{npo-Single_d1-signal-pc}
\includegraphics[width=31mm,angle=-90]{npo-Single_d1-pump-pc}
\caption{ 
\label{Fnpo-single-phase}
NPA: 
envelope-phase profiles 
for 18 and 96fs pulse durations. 
Top to bottom: 
idler, signal, pump; 
SVEA  ({--~--~--}),
GFEA 96fs ({--$\cdot$--$\cdot$--$\cdot$}),
GFEA 18fs ({------}).
}
\end{figure}

\end{subsection}

} 

%
%
%
%
%

\ifthenelse{\boolean{BoolNPD}}
{

\begin{subsection}{Non-degenerate Parametric De-amplification (NPD)}\label{idealchi2-npd}

\begin{figure}
\includegraphics[width=85mm]{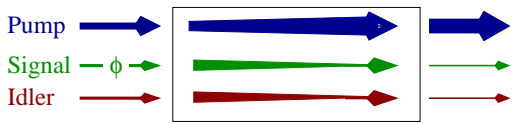}
\caption{ 
\label{Fnpd-diagram}
NPD: 
Non-degenerate Parametric De-amplification (section \ref{idealchi2-npd}). The thickness
of the arrows is intended to give an indication of how the energy
of the field components changes during propagation through the crystal.
}
\end{figure}

As a variant on the case just treated, signal and idler pulses with equal
numbers of photons were injected, and the relative phases of the pulses set to
ensure that the signal and idler experience initial de-amplification (see fig.
\ref{Fnpd-diagram}). Since the subsequent evolution is sensitive to phase
changes, and the finite pulse length terms in the GFEA affect the phases, this
is an interesting situation to examine.  In the SVEA, the signal and idler
decay away towards zero as the pulses propagate, so the SVEA output signal is
just some residual part of the input.  The GFEA evolution is different, as can
be seen from eqn. (\ref{exact-BKP-altB}) --  the finite pulse lengths alter
the phase profile of the nonlinearity, and hence change the evolution of the
pulses.  During an initial period of de-amplification, the pulses undergo a
gradual phase distortion.  Then, as the discrepancy increases,  amplification
takes over.  In a comparison of SVEA and GFEA models, the effect caused by the
phase distortion is more visible when the interaction is been strong enough
for the input component of the signal pulse to be strongly depleted, and also
is much stronger for shorter pulses

The GFEA
signal pulse profiles on exit from the crystal as a function of pulse duration
are presented in fig.  \ref{Fnpo-single-phase-cf}.  
Note that the SVEA prediction corresponds to the long-pulse limit of
the GFEA figure, but those limiting features are too small to be seen.

\ifthenelse{\boolean{BoolLong}}
{
The double peak is a result of the finite
pulse length on the nonlinear polarization -- the ``phase twist'' caused by
the few cycle terms is odd, causing a node at $\tau=0$ as the field 
is amplified -- at exactly $\tau=0$, the field continues to be de-amplified.
Note that at 18fs, for example, the output signal pulse amplitude is
larger than the input value. 
}

\ifthenelse{\boolean{BoolLong}}
{
\begin{figure}
\includegraphics[width=50mm,angle=-90]{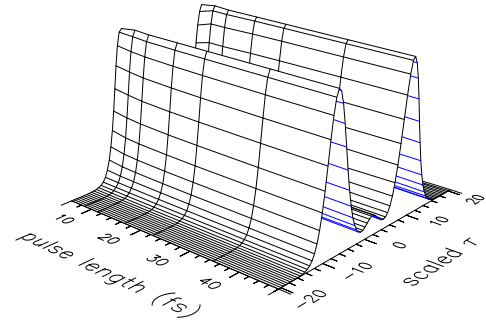}
\caption{ 
\label{Fnpo-single-phase-cf-SVEA}
NPD:
Output SVEA signal intensities for $\phi_s=\pi/4$ and $\phi_p=0$ for a range
of pulse durations: peak value $\left| A \right|^2 \approx 4 \times 10^2$).
}
\end{figure}

The SVEA profile in fig. \ref{Fnpo-single-phase-cf-SVEA} is just the residual
input signal pulse which has not yet been fully ``de-amplified''; this residue
can also be seen as a shoulder on the wings of the GFEA profiles if plotted on
a $\log$ scale.  Fig. \ref{Fnpo-single-phase-cf} shows the output GFEA signal
pulse profiles on exit from the crystal, note the enormous difference in the
few-cycle dependence of the SVEA and GFEA theories -- both in character and
amplitude.  

}

\begin{figure}
\includegraphics[width=50mm,angle=-90]{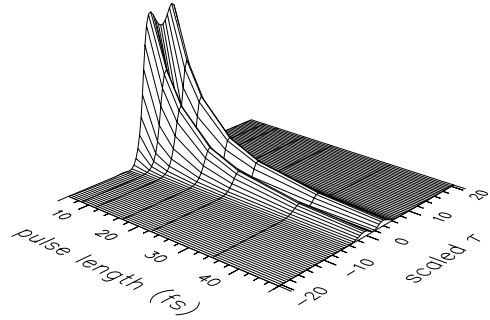}
\caption{ 
\label{Fnpo-single-phase-cf}
NPD:
Output GFEA signal intensities for $\phi_s=\pi/4$ and $\phi_p=0$ for a range
of pulse durations, peak value $\approx 6 \times 10^5$.
Equivalent SVEA results are very different: they are the same for 
all pulse lengths, are too small to show up on the scale of this graph
(being $\sim$3\% of the height of the 48fs GFEA peaks), 
and the two peaks are located further from the origin (at $\tau \approx 
\pm 10$).
}
\end{figure}

\ifthenelse{\boolean{BoolLong}}
{
\begin{figure}
\includegraphics[width=50mm,angle=-90]{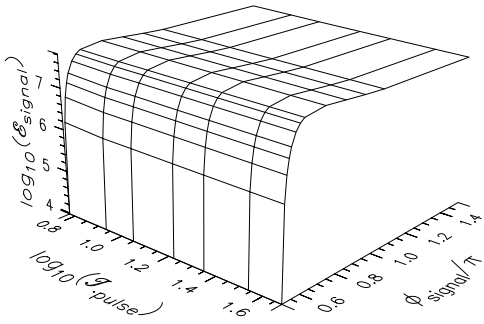}
\caption{ 
\label{Fnpo-single-phase-SVEA-ecf}
NPD:
Output SVEA pulse energies ($\mathcal{E}_{\mathrm{signal}} =
  \int{\left| A_{\mathrm{signal}} \right|^2 d\tau}$, arbitrary units) 
for a range of 
initial signal (envelope) phases $\phi_s$ 
and pulse lengths $\mathcal{T}_{\mathrm{pulse}}$.
}
\end{figure}

In fig. \ref{Fnpo-single-phase-SVEA-ecf} we see how the SVEA behaviour changes
both with pulse length and initial phase -- here we see perfect
de-amplification.

}

\begin{figure}
\includegraphics[width=70mm,angle=0]{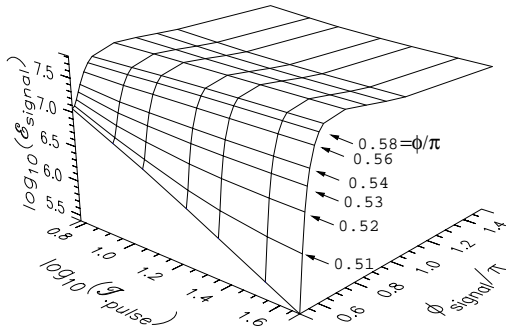}
\caption{ 
\label{Fnpo-single-phase-ecf}
NPD:
Output GFEA pulse energies ($\mathcal{E}_{\mathrm{signal}} =
  \int{\left| A_{\mathrm{signal}} \right|^2 d\tau}$, arbitrary units) 
for a range of 
initial signal (envelope) phases $\phi_s$ 
and pulse lengths $\mathcal{T}_{\mathrm{pulse}}$.
The intensities and times are scaled in our usual way. 
The SVEA results are nearly identical for all pulse lengths, and 
differ from the 48fs (i.e. $\log_{10}(48)=1.68$) results in 
that the near $\phi_{\mathrm{signal}}=\pi/2$ give significantly
lower energies -- down to $10^4$ rather than $3 \times 10^5$ for $\pi/2$.
}
\end{figure}


The GFEA output pulse energies are displayed in fig.
\ref{Fnpo-single-phase-ecf}, which shows how the behaviour changes both with
pulse length and initial phase.  The data for $\phi_{\mathrm{signal}}=\pi/2$
demonstrates the effects of exact initial conditions and finite pulse
length; maximum de-amplification occurs in the long-pulse (SVEA) limit.  If we
instead start with a signal phase slightly different from $\pi/2$, e.g.
$0.51\pi$, the de-amplification is less efficient and will eventually be
overtaken by the amplification, even for the SVEA model.   Consequently,
comparisons for imperfect initial phases are dependent on the length of the
crystal.  However, since we use a scaling procedure, the results still behave
in a systematic way, even if they are not completely generic.  

Of course, changing other initial conditions can also disturb the
de-amplification: e.g. different  numbers of signal and idler photons. 
Although both signal and idler will initially be de-amplified, as they
approach zero photon number, one field will ``overshoot'' the zero and be
inverted.  This alters the phase relationships, and so again amplification
takes over.  As an example, simulations based on our 18fs pulses suggested
that photon number mismatches of about one percent would not noticeably
disrupt the appearance of either fig.  \ref{Fnpo-single-phase-cf} or
\ref{Fnpo-single-phase-ecf}.

\ifthenelse{\boolean{BoolLong}}
{These
comparisons testing the sensitivity of these results to mismatches in the
initial conditions are of course dependent on our chosen system parameters. 
For example, either  more intense pulses or longer crystals would diminish the
difference between SVEA and GFEA predictions; conversely, weaker pulses or
shorter crystals would enhance them.
}

\end{subsection}

} 

\end{section}



\begin{section}{Optical Parametric Oscillation (OPO)}\label{opo}

We move on from optical parametric amplification to a synchronously-pumped
Optical Parametric Oscillator (OPO).   As shown in fig. \ref{Fopo-diagram}, we
considered the case of a LiNbO$_3$ crystal in an optical cavity with mirrors
that reflect the signal wavelength only.  The oscillator is driven by a train
of gaussian pump pulses whose periodicity closely matches the natural period of
the cavity, and which amplify and then sustain the signal pulse confined
within it.  The cavity length can be ``tuned'' about exact synchronisation. 
The idler pulse, generated when the signal pulse interacts with each new pump
pulse, is transmitted through the output mirror with the pump, while the
signal is strongly reflected.  For a given set of parameter values, we
modelled the development of the signal pulse over many cavity transits until
it reached a steady state. Typically, we found that the signal stabilised in
several hundred transits although, in a few cases, no equilibrium was achieved
and the system oscillated indefinitely.  \ifthenelse{\boolean{BoolLong}} { We
studied the evolution for a range of phase-matching conditions ($\Delta k = 0$
to $24 \times 10^{-3} /\mu$m, and cavity length tunings ($-16$fs to $+16$fs). 
Except where stated, the figures contain data for the perfectly phase matched
and sychronised case.  } { Here we present results for the perfectly phase
matched, sychronised case.  }

\begin{figure}
\includegraphics[width=85mm,angle=-0]{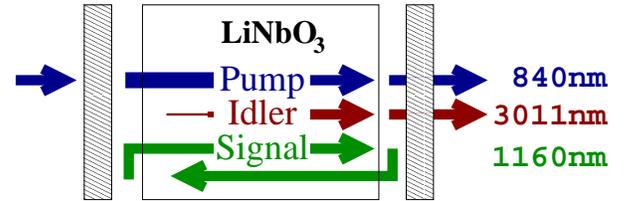}
\caption{ 
\label{Fopo-diagram}
Simplifed optical parametric oscillator experiment set-up 
(see section \ref{opo}). 
}
\end{figure}

The steady state was determined by propagating the pulses through many passes
of the oscillator until the modulus-squared
of the pulse {\em envelopes} had stabilized.

\ifthenelse{\boolean{BoolLong}}
{ 
\begin{subsection}{Dispersion Scaled OPO}\label{opo-fullscaled}

As discussed in section \ref{scaling}, it is possible to make the SVEA
propagation equation completely scale invariant if we modify the crystal 
dispersion parameters as well as the pulse 
length, energy,
and crystal length.  Although this is not an experimentally achievable goal,
it is instructive to look at what happens to the OPO output in this instance. 
To make automatic processing of the results more convenient, we adjusted 
the crystal lengths to be exact powers of two, which differed 
slightly from those derived from the standard
reference length: i.e. the 24fs pulse case related to a crystal 
length of 512$\mu$m, not
500$\mu$m.  As expected from the scaling, and
shown on figs. \ref{F-fullscale-SVEA-sq}, \ref{F-fullscale-SVEA-ph}, the
scaled SVEA results were independent of pulse length.

In contrast the GFEA pulse profiles  have a
more complex behaviour, as shown on figs. 
\ref{F-fullscale-GFEA-sq}, \ref{F-fullscale-GFEA-ph}.  For the chosen
parameters, they have a profile which becomes increasingly distorted at
shorter pulses; but there is also an abrupt transition to what is an SVEA-like
pulse profile at between 36 and 48fs.

On fig. \ref{F-fullscale-diff} we can see
this transition and the expected gradually convergence of GFEA and SVEA; the
maximum differences plotted are about 10\% of the pulse heights, this message
is repeated on fig. \ref{F-fullscale-diffisp}.  If the convergence of GFEA to
SVEA seems slower than inutitively expected,  this could be partly explained by
noting that the effect of the extra $\partial_\tau$ term caused by the non
slowly-varying nature of the pulses has less distance over which to
accumulate, so a similar discrepancy between SVEA and SEWA at e.g. 24fs and
12fs means that the 12fs case accumulates additional evolution at twice the
rate as at 24fs -- as indeed is to be expected, since broadly speaking, the
time derivative of a pulse envelope will double as its pulse length is halved.

\begin{figure}
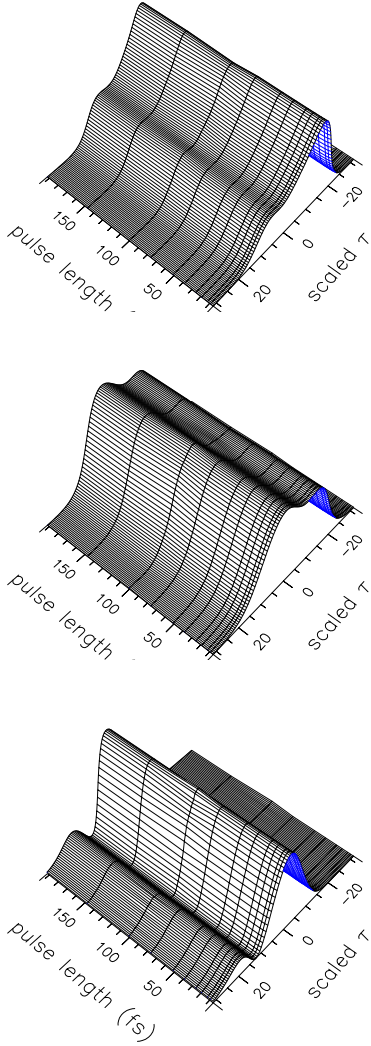

\includegraphics[width=46mm,angle=-90]{SVEA3_d2-idler-p0000s=0000pp=006fs-sq-0.epsi}
\includegraphics[width=46mm,angle=-90]{SVEA3_d2-signal-p0000s=0000pp=006fs-sq-0.epsi}
\includegraphics[width=46mm,angle=-90]{SVEA3_d2-pump-p0000s=0000pp=006fs-sq-0.epsi}
\caption{ 
\label{F-fullscale-SVEA-sq}
Dispersion  Scaled OPO: 
Time domain representation of the SVEA modulus-squared of the 
pulse envelopes, for a range of injected pump pulse durations: 
6-192fs. (bottom to top) pump, signal, and idler.
}
\end{figure}

\begin{figure}
\includegraphics[width=46mm,angle=-90]{SVEA3_d2-idler-p0000s=0000pp=006fs-ph-0.epsi}
\includegraphics[width=46mm,angle=-90]{SVEA3_d2-signal-p0000s=0000pp=006fs-ph-0.epsi}
\includegraphics[width=46mm,angle=-90]{SVEA3_d2-pump-p0000s=0000pp=006fs-ph-0.epsi}
\caption{ 
\label{F-fullscale-SVEA-ph}
Dispersion  Scaled OPO: 
Time domain representation of the SVEA amplitude phase, 
for a range of injected pump pulse durations: 
6-192fs. (bottom to top) pump, signal, and idler.
}
\end{figure}

\begin{figure}
\includegraphics[width=46mm,angle=-90]{GFEA3_d2-idler-p0000s=0000pp=006fs-sq-0.epsi}
\includegraphics[width=46mm,angle=-90]{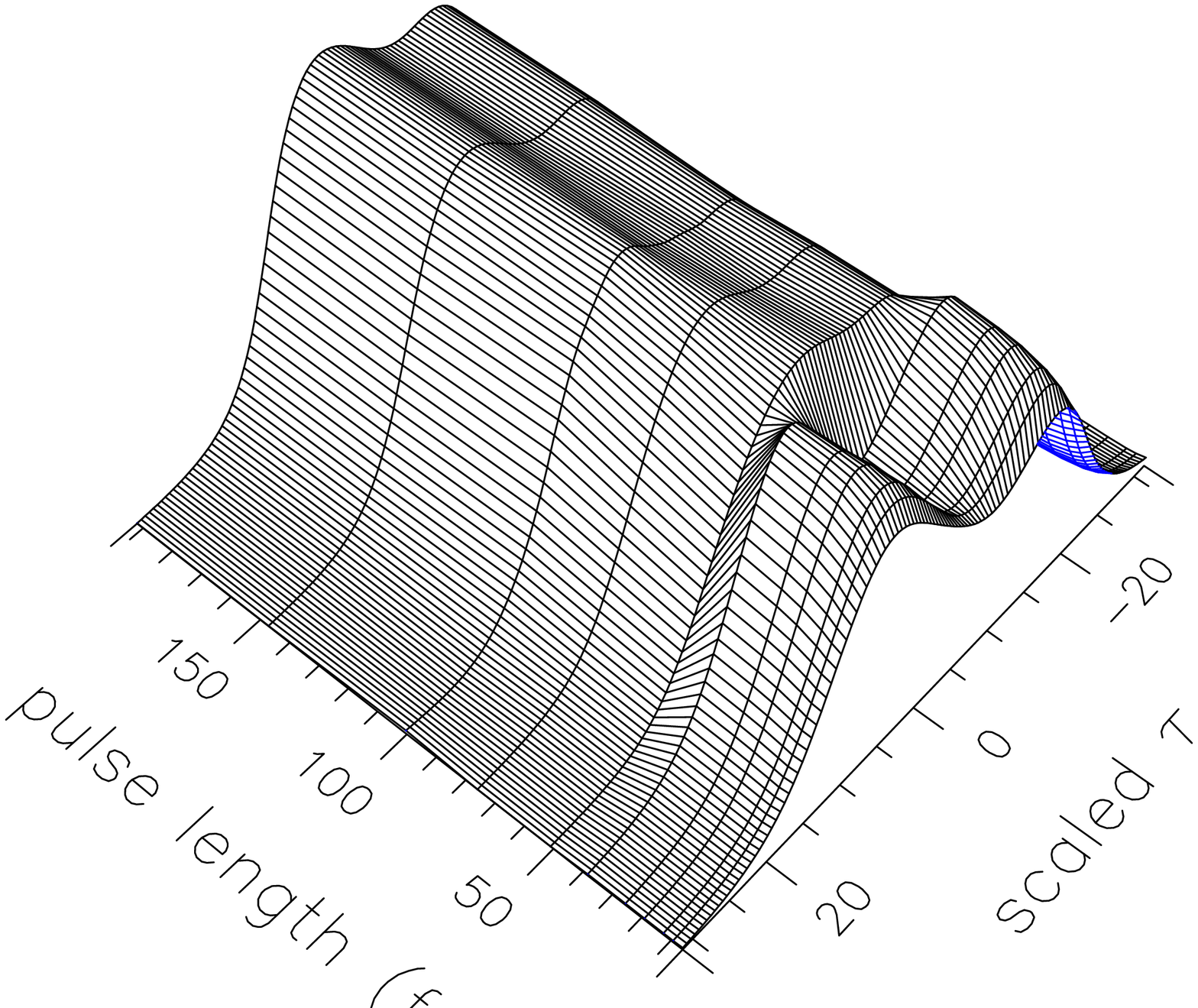}
\includegraphics[width=46mm,angle=-90]{GFEA3_d2-pump-p0000s=0000pp=006fs-sq-0.epsi}
\caption{ 
\label{F-fullscale-GFEA-sq}
Dispersion  Scaled OPO: 
Time domain representation of the GFEA modulus-squared of the 
pulse envelopes, for a range of injected pump pulse durations: 
6-192fs. (bottom to top) pump, signal, and idler.
}
\end{figure}

\begin{figure}
\includegraphics[width=46mm,angle=-90]{GFEA3_d2-idler-p0000s=0000pp=006fs-ph-0.epsi}
\includegraphics[width=46mm,angle=-90]{GFEA3_d2-signal-p0000s=0000pp=006fs-ph-0.epsi}
\includegraphics[width=46mm,angle=-90]{GFEA3_d2-pump-p0000s=0000pp=006fs-ph-0.epsi}
\caption{ 
\label{F-fullscale-GFEA-ph}
Dispersion  Scaled OPO: 
Time domain representation of the GFEA amplitude phase, 
for a range of injected pump pulse durations: 
6-192fs. (bottom to top) pump, signal, and idler.
}
\end{figure}

\begin{figure}
\includegraphics[width=46mm,angle=-90]{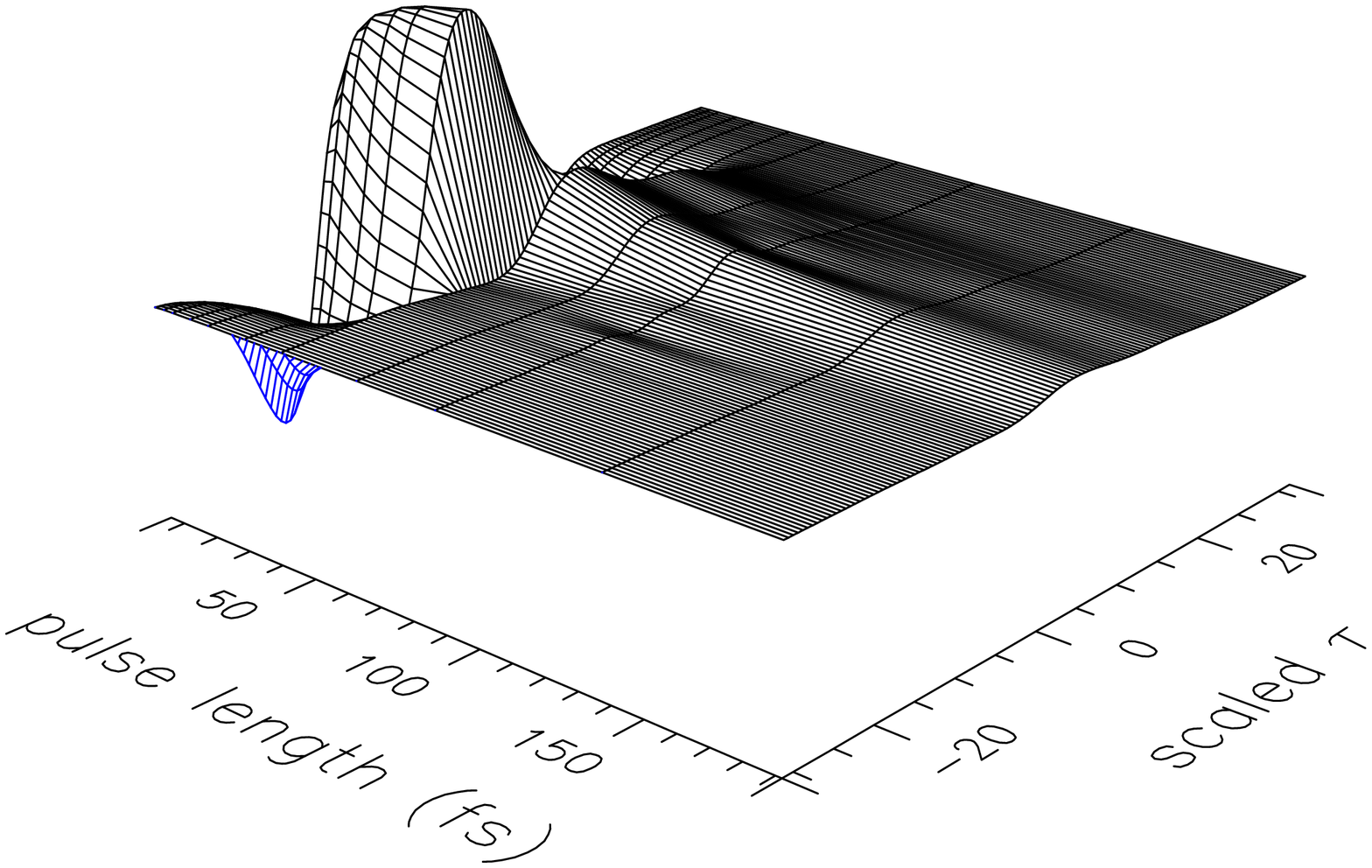}
\includegraphics[width=46mm,angle=-90]{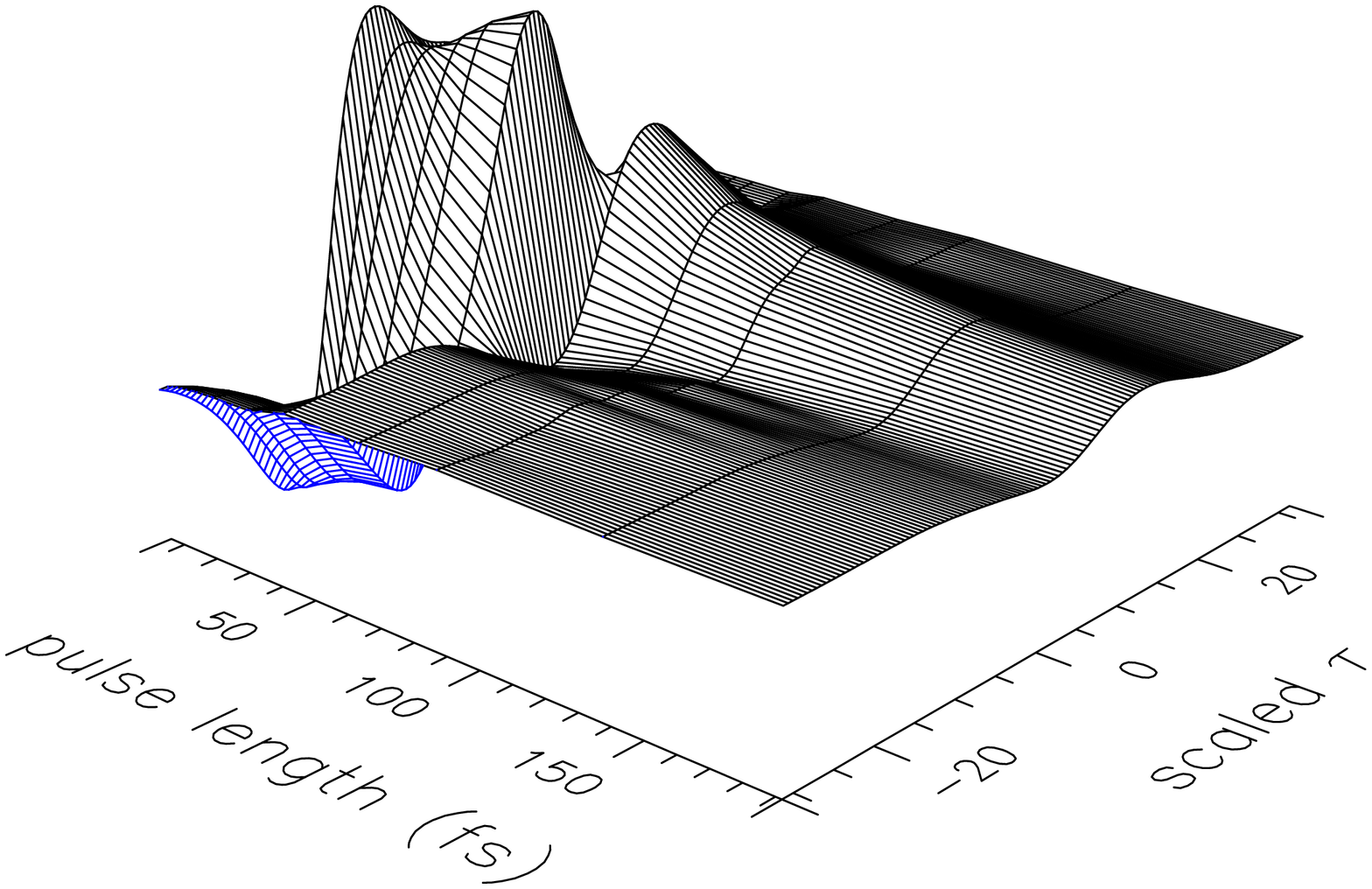}
\includegraphics[width=46mm,angle=-90]{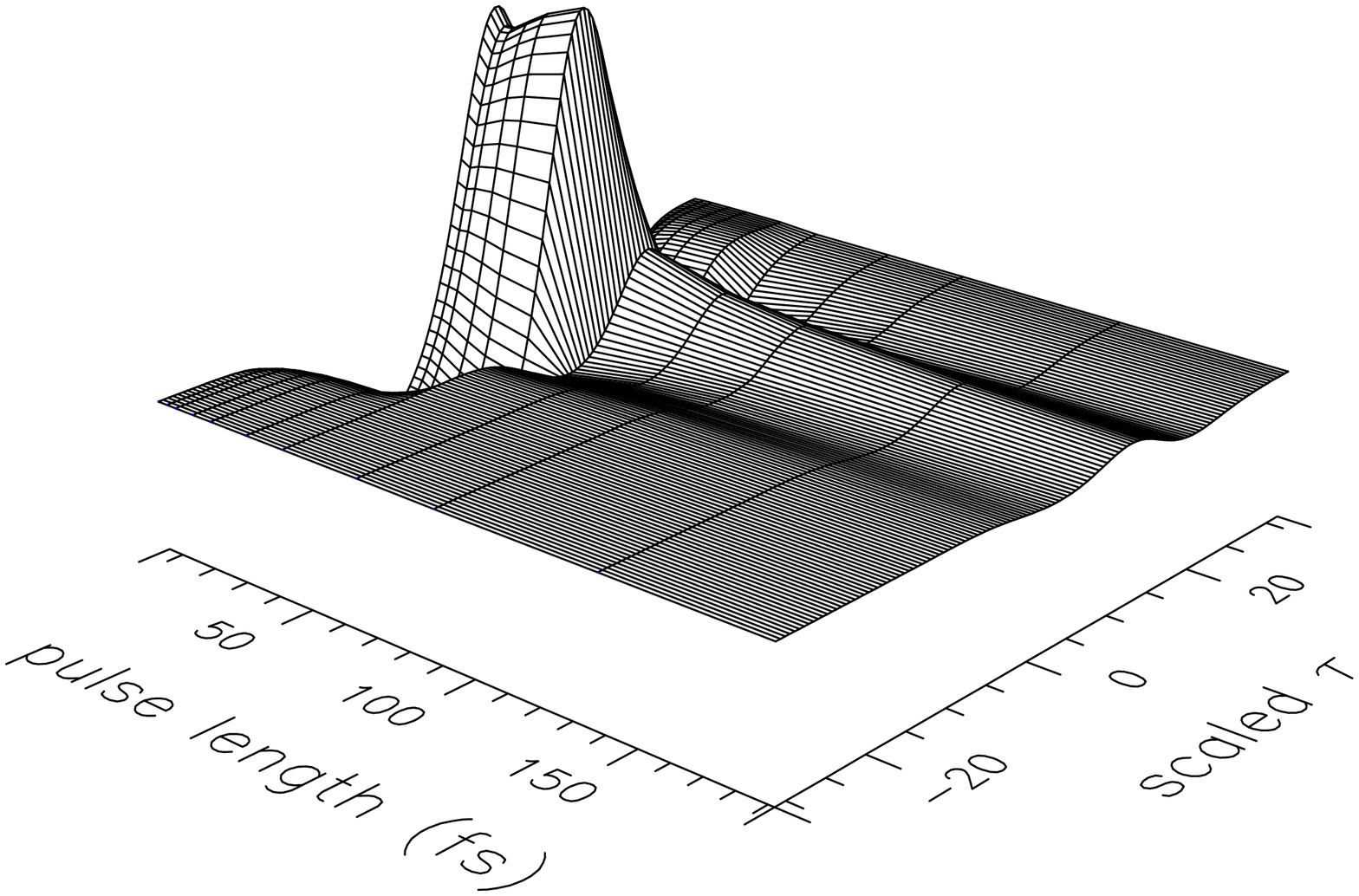}
\caption{ 
\label{F-fullscale-diff}
Dispersion  Scaled OPO: 
Time domain representation of the difference between the modulus-squared 
of the pulse envelopes between SVEA and GFEA theories, for a range of 
injected pump pulse durations: 
6-192fs. (bottom to top) pump, signal, and idler.
}
\end{figure}

\begin{figure}
\includegraphics[width=41mm,angle=-90]{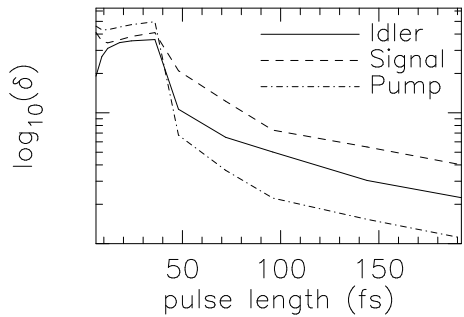}
\caption{ 
\label{F-fullscale-diffisp}
Dispersion Scaled OPO: 
Maximum difference $\delta$ between GFEA and SVEA simulations over the 
middle quarter of the scaled $\tau$ range, on a $\log_{10}$ scale.
}
\end{figure}

Note: the slight disruption to the wings of the 92fs results are because 
the time width allowed for by the simulation is smaller when 
compared to the pulse width for this particular case.

\end{subsection}

}

\begin{subsection}{Scaled OPO}\label{opo-scaled}

The complex nature of the dynamics, which arises from repetitive cycling of
the signal pulse in the presence of many interacting processes 
makes the isolation and analysis of few-cycle effects within the different
models quite complicated.  Fig. \ref{FtimeA2} shows intensity profiles
for the pump, signal, and idler (bottom to top in each frame) for the 
SVEA (dashed line) and GFEA (solid line) for four different pulse 
durations. 
\ifthenelse{\boolean{BoolOPOall}}
{Simulations with pump pulse durations of 48fs and over never 
reached a steady state, and so are not included here.  
}

The first point to note in fig. \ref{FtimeA2} is that the SVEA results are not
identical in all frames, even though the scaling procedure in section
\ref{scaling} is designed to make them, as far as possible, {\em independent}
of pulse duration.  The reason is that, as noted in section \ref{scaling}, the
dispersion scales in a different way to the group time delay, and so is not
correctly compensated by eqn. (\ref{scalingsummary}).  

\ifthenelse{\boolean{BoolOPOall}}
{
\begin{figure}
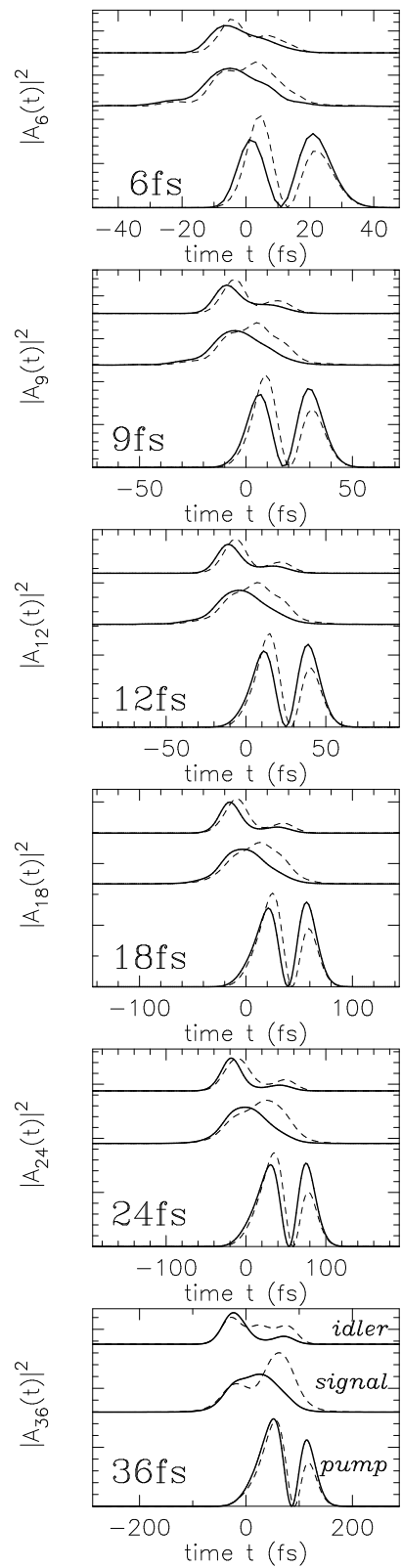

\includegraphics[width=35mm,angle=-90]{SVEA-GFCE_p-0125u-p0000s-0000pp-06fs-ix-n9-lastcfA2}
\includegraphics[width=35mm,angle=-90]{SVEA-GFCE_p-0187u-p0000s-0000pp-09fs-ix-n9-lastcfA2}
\includegraphics[width=35mm,angle=-90]{SVEA-GFCE_p-0250u-p0000s-0000pp-12fs-ix-n9-lastcfA2}
\includegraphics[width=35mm,angle=-90]{SVEA-GFCE_p-0375u-p0000s-0000pp-18fs-ix-n9-lastcfA2}
\includegraphics[width=35mm,angle=-90]{SVEA-GFCE_p-0500u-p0000s-0000pp-24fs-ix-n9-lastcfA2}
\includegraphics[width=35mm,angle=-90]{SVEA-GFCE_p-0750u-p0000s-0000pp-36fs-ix-n9-lastcfA2}
\caption{ 
\label{FtimeA2}
Scaled OPO: 
Time domain representation of the modulus-squared of the 
pulse envelopes, for a range of injected pump pulse durations: 
6fs (top), 9, 12, 18, 24, 36fs (bottom).  
For each sub-figure, the 
curves compare (bottom to top) pump, signal, and idler for the
SVEA simulations ({--~--~--})
and GFEA ones ({\bf------}).
}
\end{figure}
}
{
\begin{figure}
\includegraphics[width=35mm,angle=-90]{SVEA-GFCE_p-0125u-p0000s-0000pp-06fs-ix-n9-lastcfA2}
\includegraphics[width=35mm,angle=-90]{SVEA-GFCE_p-0250u-p0000s-0000pp-12fs-ix-n9-lastcfA2}
\includegraphics[width=35mm,angle=-90]{SVEA-GFCE_p-0500u-p0000s-0000pp-24fs-ix-n9-lastcfA2}
\includegraphics[width=35mm,angle=-90]{SVEA-GFCE_p-0750u-p0000s-0000pp-36fs-ix-n9-lastcfA2}
\caption{ 
\label{FtimeA2}
Scaled OPO: 
Time domain representation of the modulus-squared of the 
pulse envelopes, for a range of injected pump pulse durations: 
6fs (top), 12, 24, 36fs (bottom).  
For each sub-figure, the 
curves compare (bottom to top) pump, signal, and idler for the
SVEA simulations ({--~--~--})
and GFEA ones ({\bf------}).
}
\end{figure}
}

\ifthenelse{\boolean{BoolOPOall}}
{
\begin{figure}
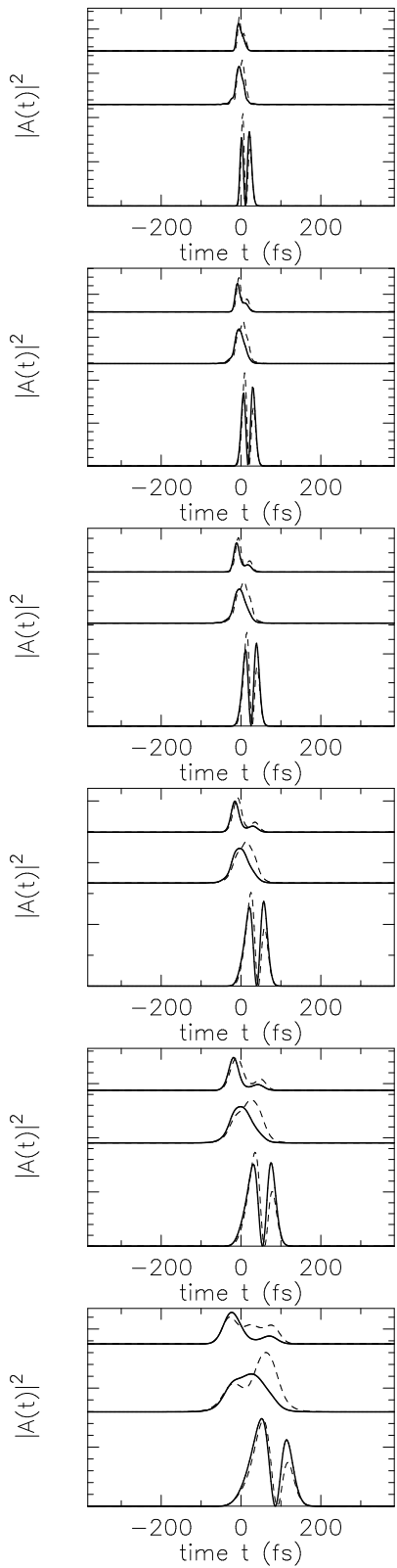

\includegraphics[width=35mm,angle=-90]{SVEA-GFCE_u-0125u-p0000s-0000pp-06fs-ix-n9-lastcfA2}
\includegraphics[width=35mm,angle=-90]{SVEA-GFCE_u-0187u-p0000s-0000pp-09fs-ix-n9-lastcfA2}
\includegraphics[width=35mm,angle=-90]{SVEA-GFCE_u-0250u-p0000s-0000pp-12fs-ix-n9-lastcfA2}
\includegraphics[width=35mm,angle=-90]{SVEA-GFCE_u-0375u-p0000s-0000pp-18fs-ix-n9-lastcfA2}
\includegraphics[width=35mm,angle=-90]{SVEA-GFCE_u-0500u-p0000s-0000pp-24fs-ix-n9-lastcfA2}
\includegraphics[width=35mm,angle=-90]{SVEA-GFCE_u-0750u-p0000s-0000pp-36fs-ix-n9-lastcfA2}
\caption{ 
\label{FtimeA2x}
Scaled OPO:
As above in fig. \ref{FtimeA2}, for
6fs (top), 9, 12, 18, 24, 36fs (bottom);
but using the same time window width for all pulse widths.
}
\end{figure}
}
{}

A second rather surprising feature is that we might expect the GFEA results to
tend to the SVEA as pulse length increases, but this is not evident from the
graphs.  The explanation for this is that the steady state of the OPO can
change suddenly as the parameters are varied.  This property is highlighted in
fig. \ref{F-fullscale-GFEA-sq}, which shows the GFEA signal pulse profile for
pulse durations from 6fs to 192fs; the sudden adjustment of the GFEA when
moving from 36fs to 48fs takes it close to the SVEA, and the difference
between the two gradually disappears as the pulse duration is increased
further (see Fig. \ref{F-fullscale-diffisp} 
\ifthenelse{\boolean{BoolLong}}{)}
{and \cite{Kinsler-N-2002longFCOPO})}.  Note that the scaling procedure used
for fig. \ref{F-fullscale-GFEA-sq} is an extension of eqn.
(\ref{scalingsummary}) in that the dispersion term is also scaled, making the
SVEA results completely independent of pulse duration.

\ifthenelse{\boolean{BoolLong}}
{ 
}
{
\begin{figure}
\includegraphics[width=61mm,angle=-90]{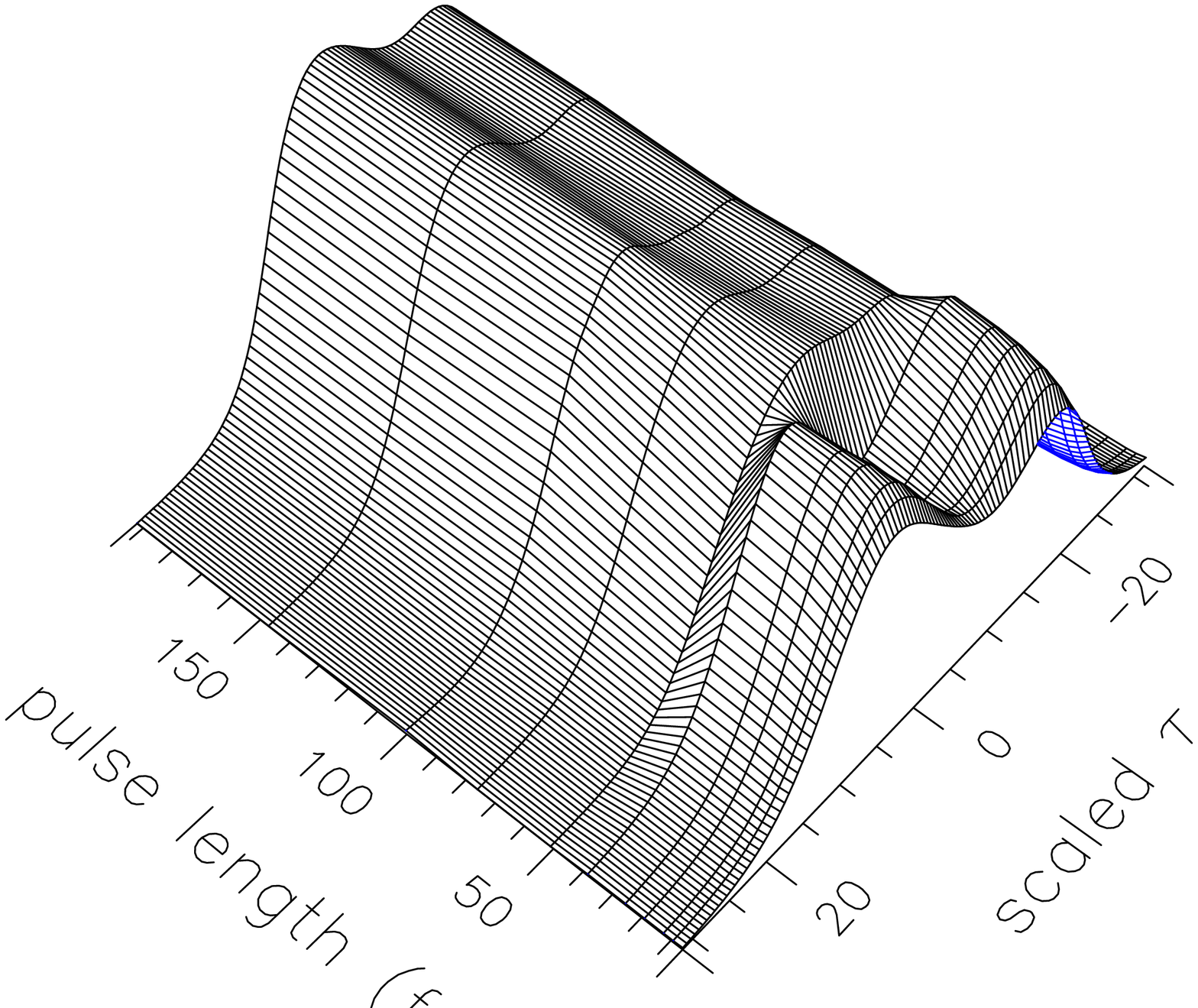}
\caption{ 
\label{F-fullscale-GFEA-sq}
Dispersion  Scaled OPO: 
Time domain representation of the GFEA signal amplitude, for a range 
injected pump pulse durations from 
6-192fs. Here the crystal dispersion is adjusted in 
addition to the other scalings to make a SVEA theory fully scale invariant.  
The SVEA profile in this case very similar to the 192fs 
GFEA profile. 
}
\end{figure}

\begin{figure}
\includegraphics[width=41mm,angle=-90]{GFEA3_d1-diff-isp-sq-Log}
\caption{ 
\label{F-fullscale-diffisp}
Dispersion Scaled OPO: 
Maximum difference between GFEA and SVEA simulations over the 
middle quarter of the scaled $\tau$ range, on a $\log_{10}$ scale.
}
\end{figure}

}

Spectral profiles corresponding to the temporal profiles of fig.
\ref{FtimeA2} \ifthenelse{\boolean{BoolOPOall}}{ and \ref{FtimeA2x}}{}
are shown in fig. \ref{FfreqA2}.  
The spectral shape for each
field is similar across all pulse durations, with a pulse of double the (time)
width naturally having half the bandwidth. Notice that the pump and signal 
spectra in the 6fs frame are close to overlapping, which indicates that the 
separation of the total EM field into distinct pump, signal, and idler 
components 
is becoming a questionable assumption.

\ifthenelse{\boolean{BoolOPOall}}
{
\begin{figure}
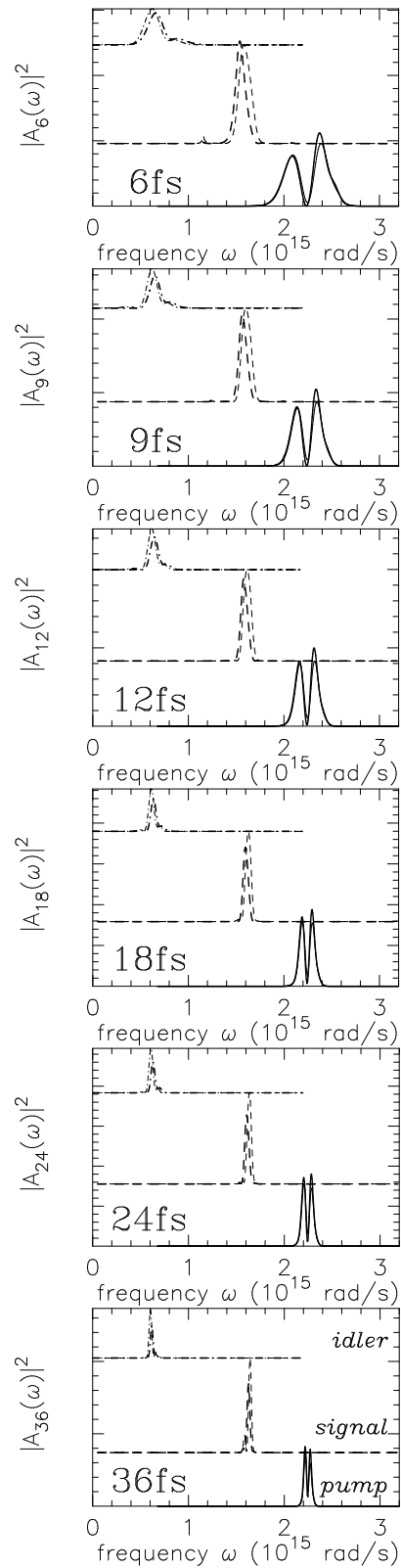

\includegraphics[width=35mm,angle=-90]{SVEA-GFCE_p-0125u-p0000s-0000pp-06fs-ix-n9-laswcfA2}
\includegraphics[width=35mm,angle=-90]{SVEA-GFCE_p-0187u-p0000s-0000pp-09fs-ix-n9-laswcfA2}
\includegraphics[width=35mm,angle=-90]{SVEA-GFCE_p-0250u-p0000s-0000pp-12fs-ix-n9-laswcfA2}
\includegraphics[width=35mm,angle=-90]{SVEA-GFCE_p-0375u-p0000s-0000pp-18fs-ix-n9-laswcfA2}
\includegraphics[width=35mm,angle=-90]{SVEA-GFCE_p-0500u-p0000s-0000pp-24fs-ix-n9-laswcfA2}
\includegraphics[width=35mm,angle=-90]{SVEA-GFCE_p-0750u-p0000s-0000pp-36fs-ix-n9-laswcfA2}
\caption{ 
\label{FfreqA2}
Scaled OPO: 
Frequency domain representation of the modulus-squared of the 
pulse envelopes, 
for pump pulse durations of 6, 12, 18, 24, and 36fs. 
For each sub-figure, the 
curves compare (bottom to top) pump, signal, and idler for the
SVEA simulations ({--~--~--})
and GFEA ones ({\bf------}).
}
\end{figure}
}
{
\begin{figure}
\includegraphics[width=35mm,angle=-90]{SVEA-GFCE_p-0125u-p0000s-0000pp-06fs-ix-n9-laswcfA2}
\includegraphics[width=35mm,angle=-90]{SVEA-GFCE_p-0250u-p0000s-0000pp-12fs-ix-n9-laswcfA2}
\includegraphics[width=35mm,angle=-90]{SVEA-GFCE_p-0750u-p0000s-0000pp-36fs-ix-n9-laswcfA2}
\caption{ 
\label{FfreqA2}
Scaled OPO: 
Frequency domain representation of the modulus-squared of the 
pulse envelopes, 
for pump pulse durations of 6, 12, and 36fs. 
For each sub-figure, the 
curves compare (bottom to top) pump, signal, and idler for the
SVEA simulations ({--~--~--})
and GFEA ones ({\bf------}).
}
\end{figure}
}

Inclusion of the carrier wave in the results raises some quite subtle issues
that need careful consideration.  It must be stressed again that the carrier
drops out of the analysis leading to eqn. (\ref{exact-BKP}). The envelope
description is therefore complete, although the phases of two of the three
envelope functions can be changed by arbitrary constants without any effect on
the computations apart from an appropriate adjustment in the phase of the
third envelope.  For instance, if the phases of the pump and signal envelopes
are changed by $\Delta \phi_p$ and $\Delta \phi_s$, the phase of the idler
envelope is changed by $\Delta \phi_i = \Delta \phi_p - \Delta \phi_s$. 
Adjustments of this kind show up in the results only if graphs of the complete
electric field profiles, including the carrier waves, are displayed, as in
fig. \ref{FtimeEr}.  If the simulations in that figure were re-run with
differing envelope phases, this would be reflected in temporal displacements
of the carrier-like oscillations beneath the envelopes.

A further interesting feature is that, while the moduli of the pulse enveleopes
may have stabilised in a simulation, the envelope phases can (and usually do)
change from pass to pass; this process continues indefinitely, so a movie made
up of frames from successive transits would show the pump, signal, and idler
electric field oscillations drifting across underneath the respective steady
envelope profiles.  

Note that although it is common
to talk of ``shifts in carrier phase'', in the theory {\em the carrier phase 
is specified by the initial conditions, and does not change}.  The effects
usually ascribed to ``shifts in carrier phase'' instead manifest themselves 
as a shift in the phase of the complex envelope. 

\ifthenelse{\boolean{BoolOPOall}}
{
\begin{figure}
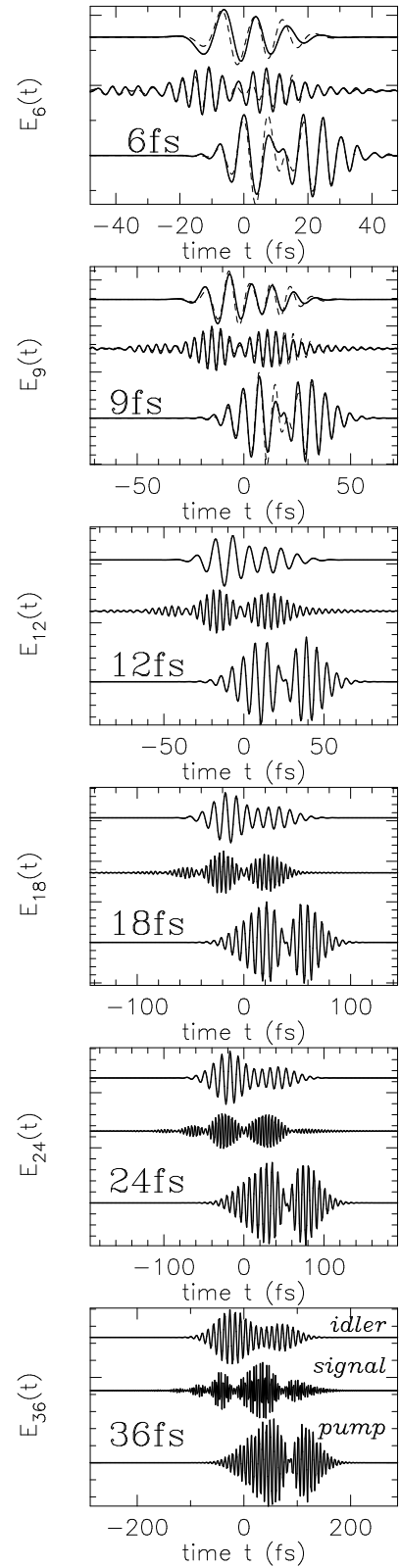

\includegraphics[width=35mm,angle=-90]{SVEA-GFCE_p-0125u-p0000s-0000pp-06fs-ix-n9-lastcfEr}
\includegraphics[width=35mm,angle=-90]{SVEA-GFCE_p-0187u-p0000s-0000pp-09fs-ix-n9-lastcfEr}
\includegraphics[width=35mm,angle=-90]{SVEA-GFCE_p-0250u-p0000s-0000pp-12fs-ix-n9-lastcfEr}
\includegraphics[width=35mm,angle=-90]{SVEA-GFCE_p-0375u-p0000s-0000pp-18fs-ix-n9-lastcfEr}
\includegraphics[width=35mm,angle=-90]{SVEA-GFCE_p-0500u-p0000s-0000pp-24fs-ix-n9-lastcfEr}
\includegraphics[width=35mm,angle=-90]{SVEA-GFCE_p-0750u-p0000s-0000pp-36fs-ix-n9-lastcfEr}
\caption{ 
\label{FtimeEr}
Scaled OPO: 
Time domain representation of the electric fields of the pulse, 
for pump pulse durations of 6, 12, 18, 14, and 36fs. 
For each sub-figure, the solid 
curves ({\bf------}) compare (bottom to top) pump, signal, and idler for the
GFEA simulations, for 6fs and 9fs the SVEA fields are also indicated 
 ({--~--~--}).
The phases are chosen so that the maximum excursion of the 
signal envelope is purely real valued, and the idler phase is chosen
so that $\phi_s+\phi_i=\phi_p$.
}
\end{figure}
}
{
\begin{figure}
\includegraphics[width=35mm,angle=-90]{SVEA-GFCE_p-0125u-p0000s-0000pp-06fs-ix-n9-lastcfEr}
\includegraphics[width=35mm,angle=-90]{SVEA-GFCE_p-0250u-p0000s-0000pp-12fs-ix-n9-lastcfEr}
\includegraphics[width=35mm,angle=-90]{SVEA-GFCE_p-0750u-p0000s-0000pp-36fs-ix-n9-lastcfEr}
\caption{ 
\label{FtimeEr}
Scaled OPO: 
Time domain representation of the electric fields of the pulse, 
for pump pulse durations of 6, 12, and 36fs. 
For each sub-figure, the solid 
curves ({\bf------}) compare (bottom to top) pump, signal, and idler for the
GFEA simulations, for 6fs the SVEA fields are also indicated 
 ({--~--~--}).
The phases are chosen so that the maximum excursion of the 
signal envelope is purely real valued, and the idler phase is chosen
so that $\phi_s+\phi_i=\phi_p$.
}
\end{figure}
}

The different models discussed in this paper give significantly
different results for the pass-to-pass phase drift.  Fig.
\ref{Fphaseincrement} shows the phase change for the signal pulses as a
function of pulse length for the SVEA, SEWA, and GFEA; note that the SEWA and
GFEA results are similar to each other, while the (less accurate) SVEA
exhibits a very different dependence.  

The reference point used in calculating the phase drift is at the maximum
amplitude of the envelope of the signal pulse, which is in fact not
necessarily at the point of maximum electric field.  This is a good choice for
our purposes because it does not move between passes once a steady state is
established.  Although these phase drifts are quite small, discrepancies
between the SVEA and GFEA will quickly accumulate.

Analysis shows that the phase of the envelope at this reference changes by a
fixed amount from one pass to the next. Since this phase drift tends to be
small, a slow evolution of the E field profile can be seen between passes in
any given simulation.  Further, the accumulated difference between the less
accurate SVEA prediction and the GFEA can lead to large differences in the
predicted electric field profiles, even if the envelope profiles happened to
be similar.
Note that the idler pulse envelope phase drift is
in the opposite direction to the signal drift, since the pump pulse arrives 
with the same envelope phase at the beginning of each pass.

The details of fig. \ref{Fphaseincrement} are less important than its
message.  Getting an output of either an identical idler pulse from pass to
pass or at least a well understood pulse shape progression can be important, 
and this will be achieved using accurate models or by experiment based
on accurate models.  We see that even for relatively long pulses the SVEA
theory predicts a different phase drift to that of the SEWA/GFEA theory;
and that the trend is different  even when using a sequence of parameters
designed to minimise the differences between simulations of different pulse
lengths.

\begin{figure}
\includegraphics[width=35mm,angle=-90]{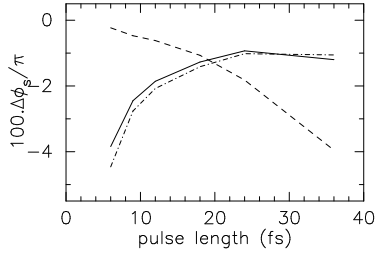}
\caption{ 
\label{Fphaseincrement}
Scaled OPO: 
Pass-to-pass phase drift for a range of parameters, 
comparing 
SVEA ({--~--~--}),
GFEA ({-----}), 
and SEWA ({--$\cdot$--$\cdot$--$\cdot$}) 
simulation results.  The differences are taken between the
phase at the peak of the modulus-squared of the envelopes
at the end of one pass of the signal pulse and the next.
}
\end{figure}

\end{subsection}


%
%
%
%
%

\ifthenelse{\boolean{BoolOPOsync}}
{

\begin{subsection}{Synchronisation and Phase Mismatch}\label{opo-sync}

The phase drift varies as a function of the phase mismatch and 
the cavity synchronisation: the data is shown in Figs.
\ref{Foposyncpp-GFEA}, \ref{Foposyncpp-SVEA}.  The glitches in the data 
are caused by occasional difficulties in evaluating the phase of the 
envelopes and the wide variations that can occur for near-zero fields --
not all the data points represent good steady states.

\begin{figure}
\includegraphics[width=43mm,angle=-90]{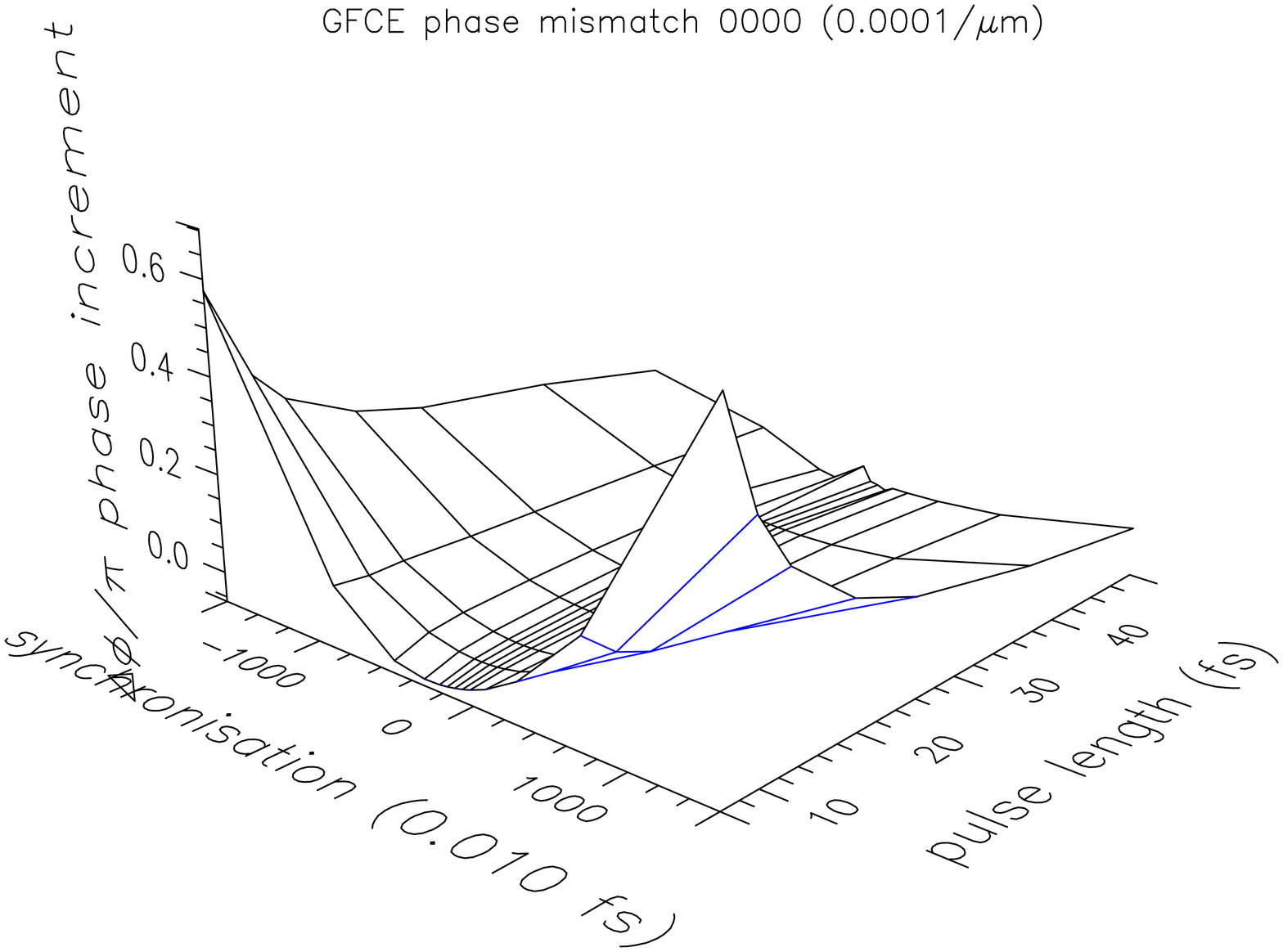}
\includegraphics[width=43mm,angle=-90]{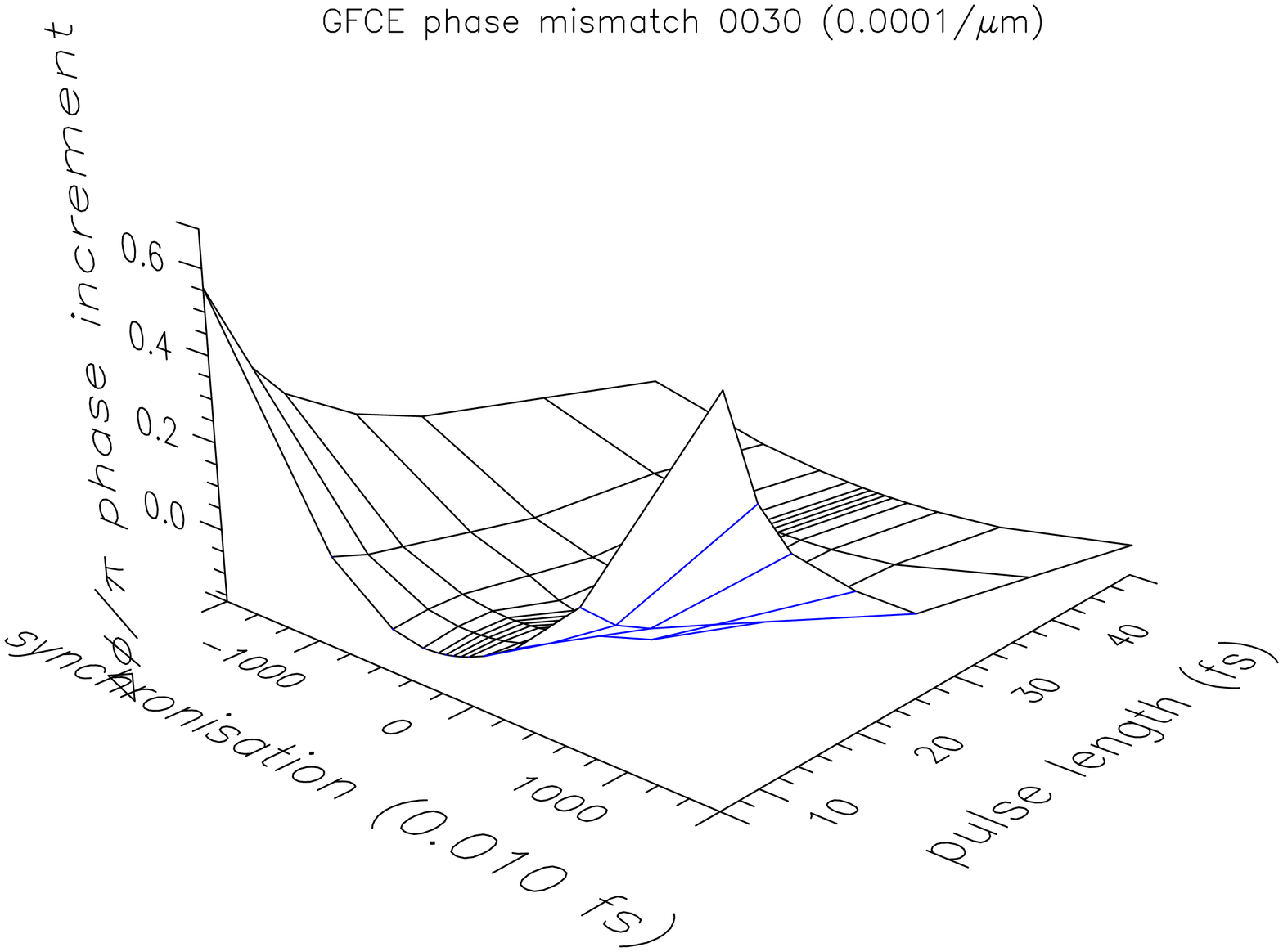}
\includegraphics[width=43mm,angle=-90]{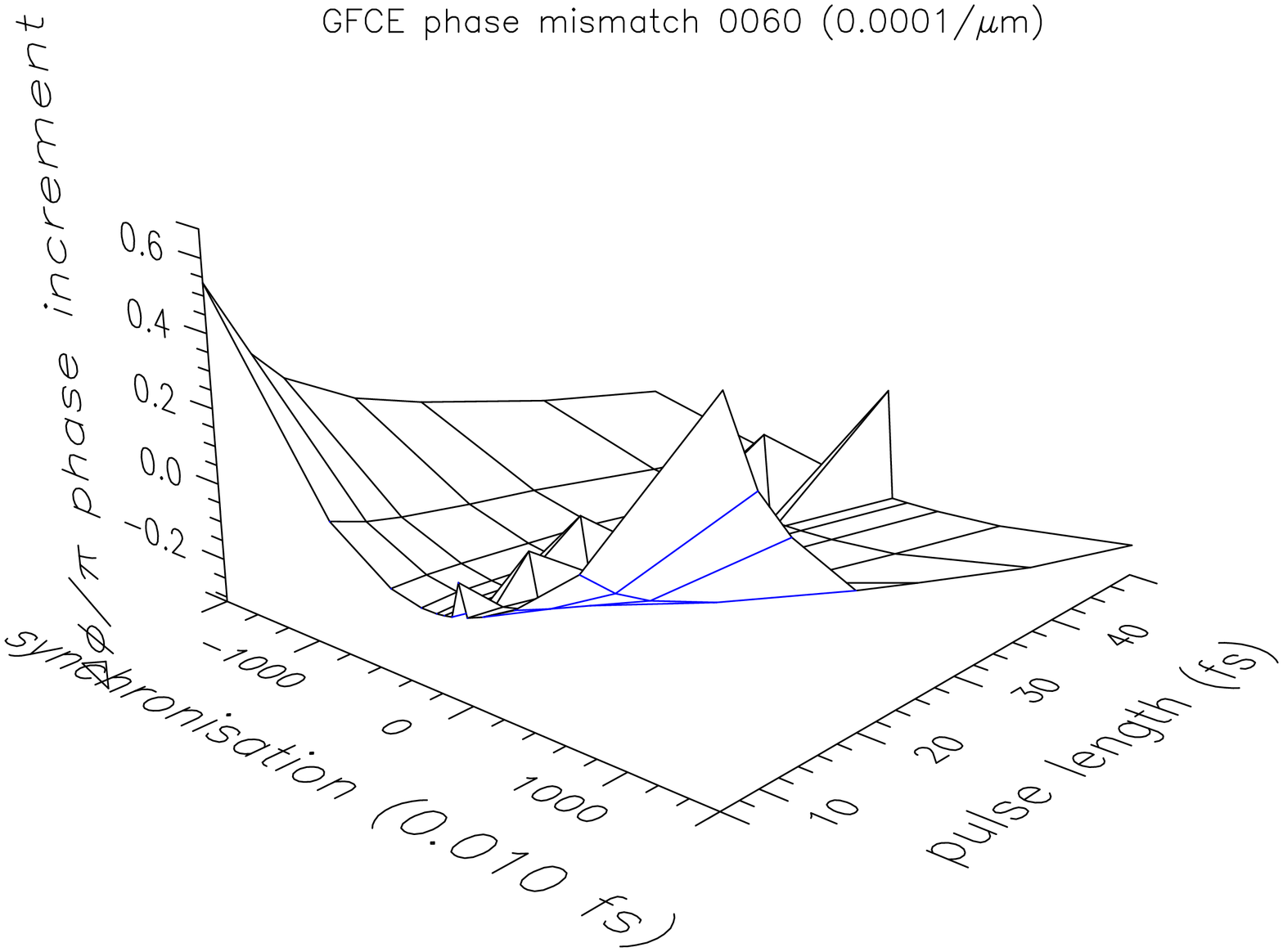}
\includegraphics[width=43mm,angle=-90]{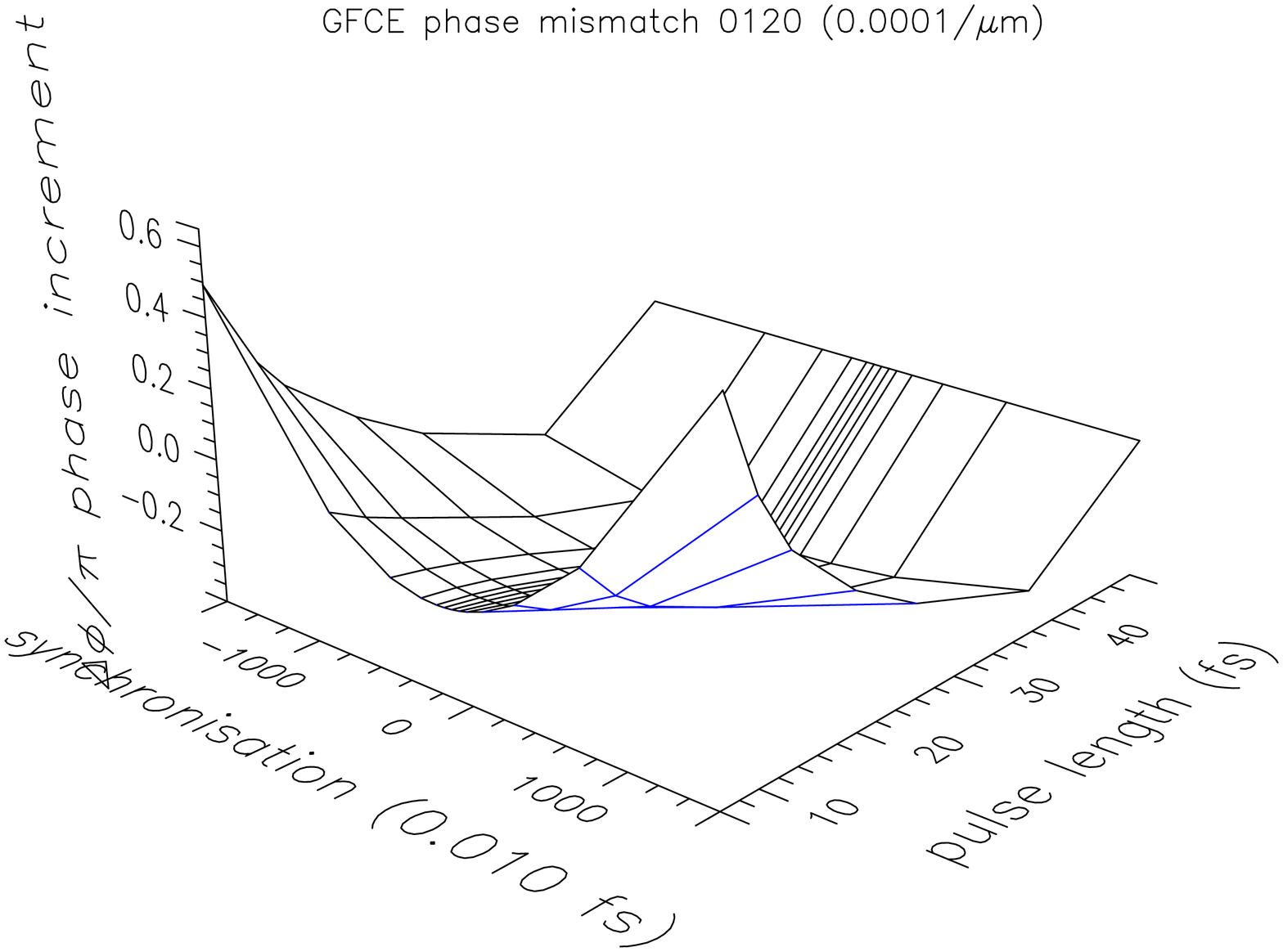}
\includegraphics[width=43mm,angle=-90]{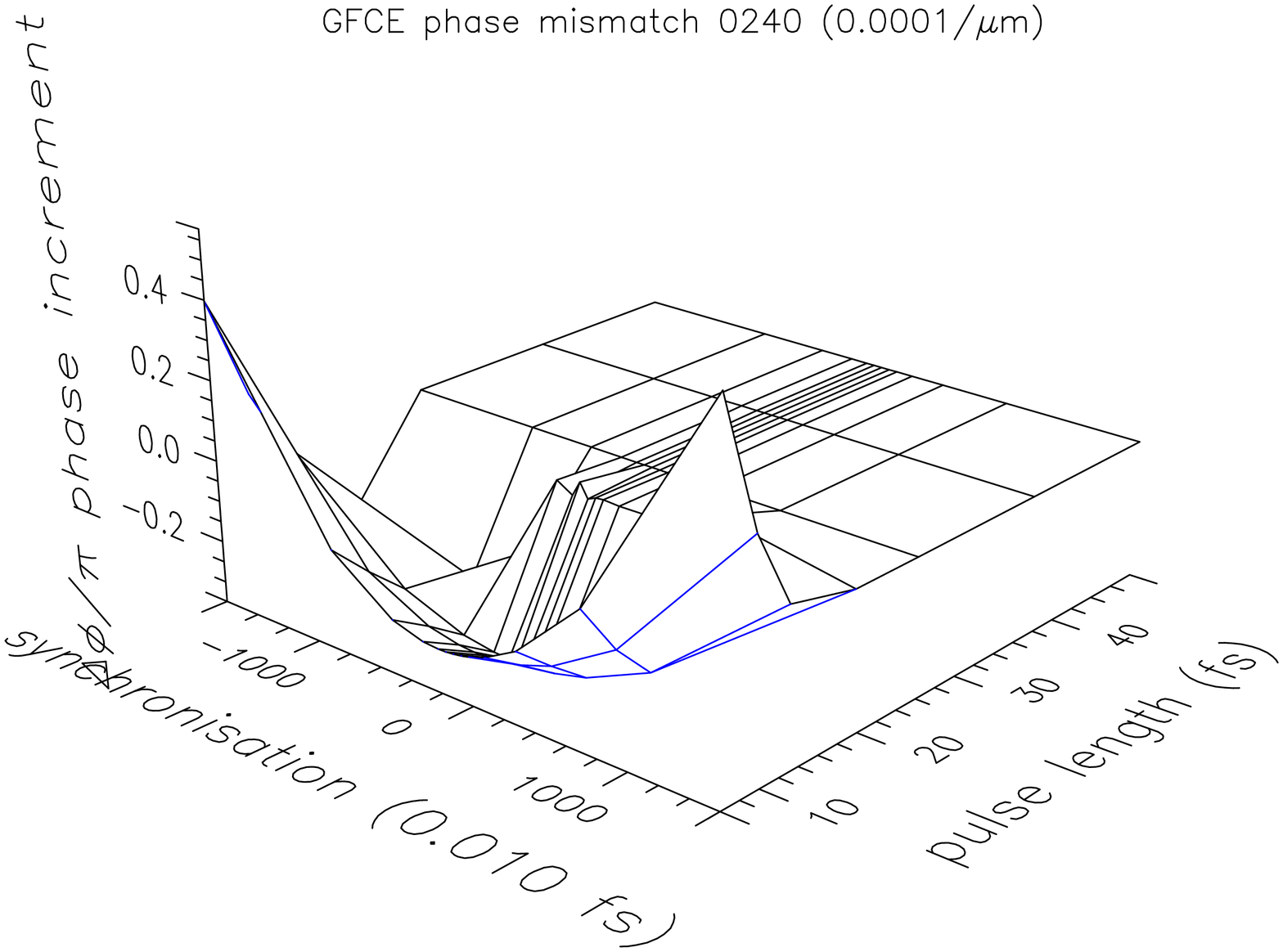}
\caption{
\label{Foposyncpp-GFEA}
Scaled OPO: GFEA phase mismatch vs sync 
} 
\end{figure}

\begin{figure}
\includegraphics[width=43mm,angle=-90]{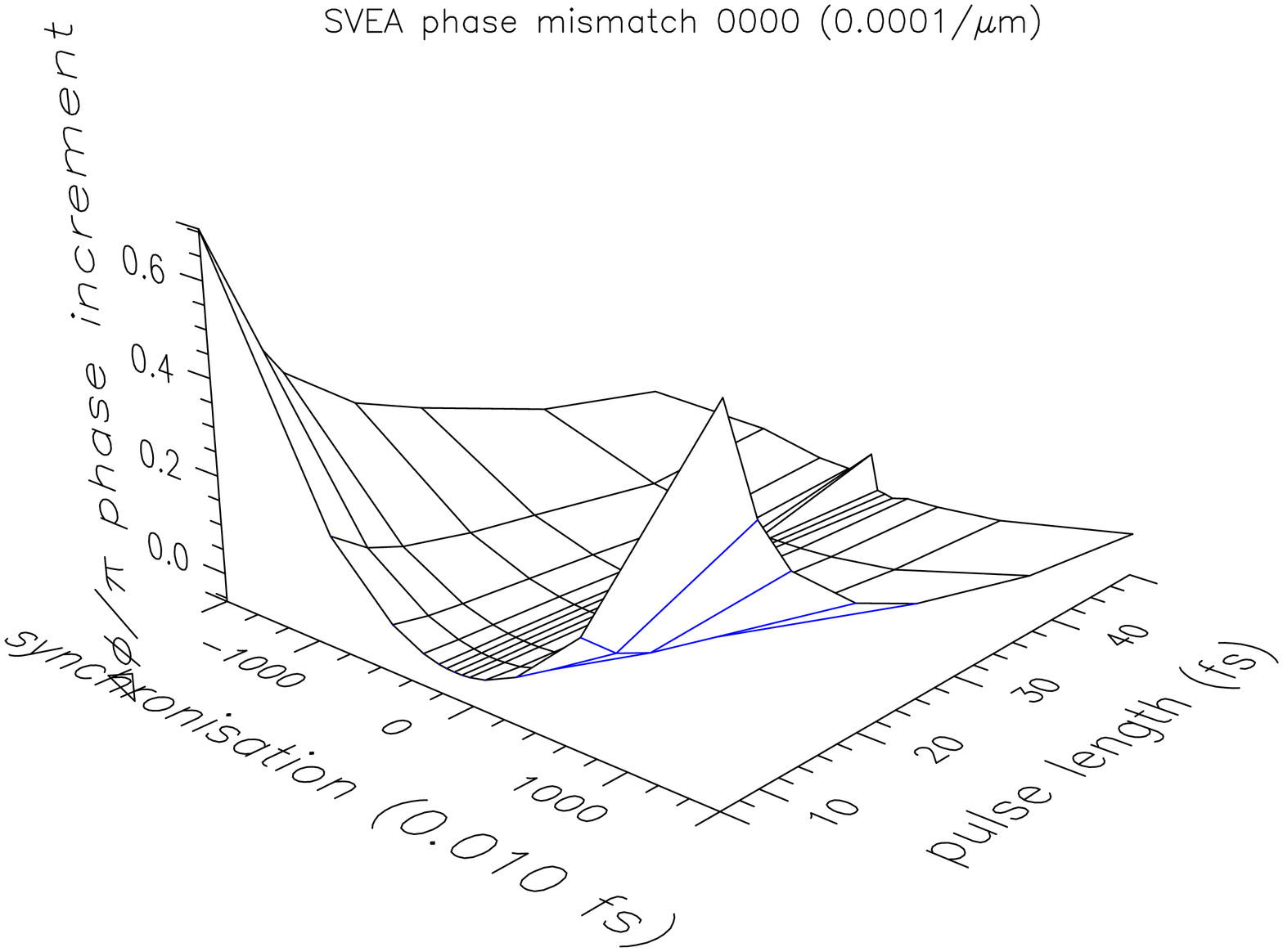}
\includegraphics[width=43mm,angle=-90]{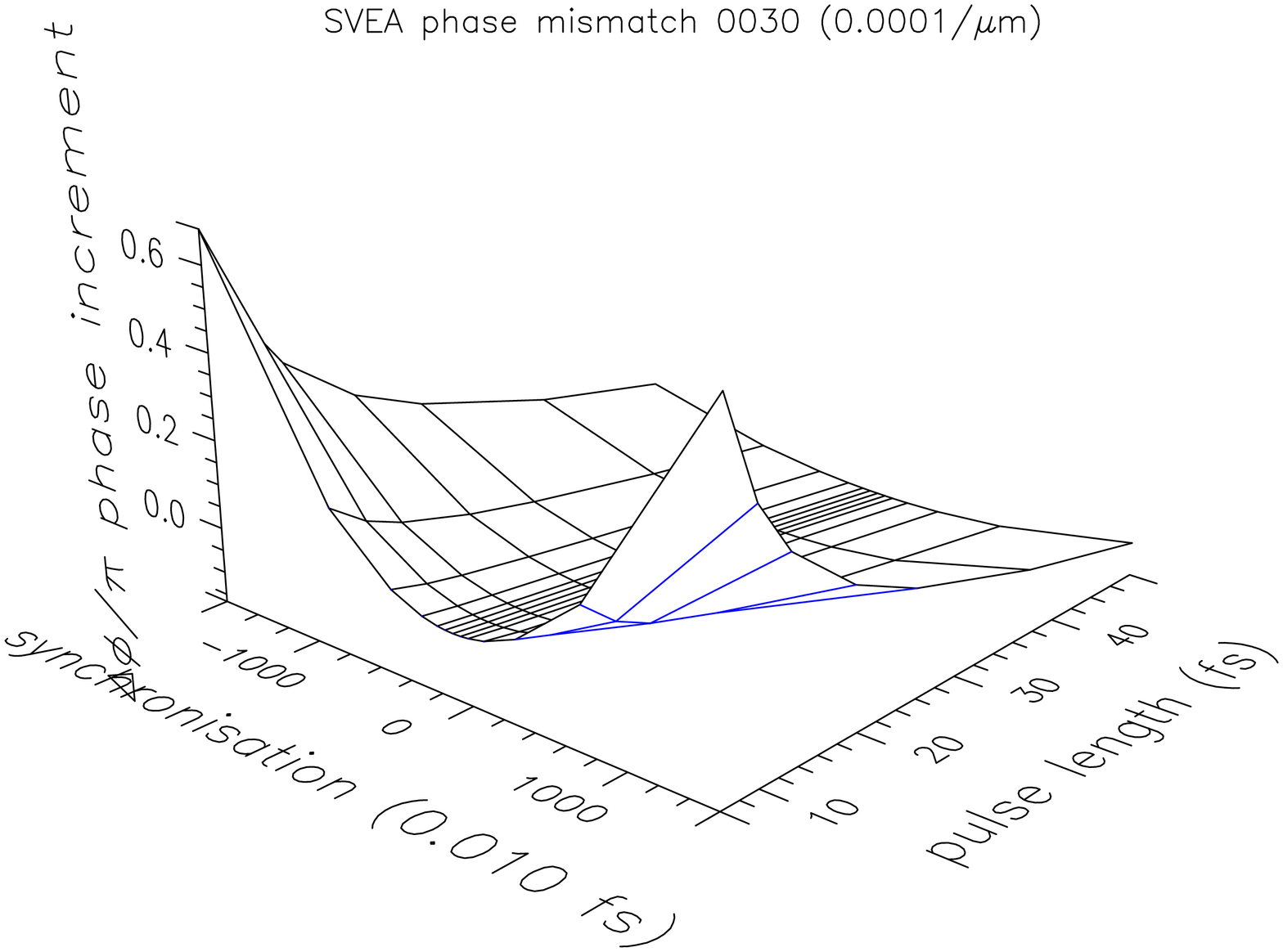}
\includegraphics[width=43mm,angle=-90]{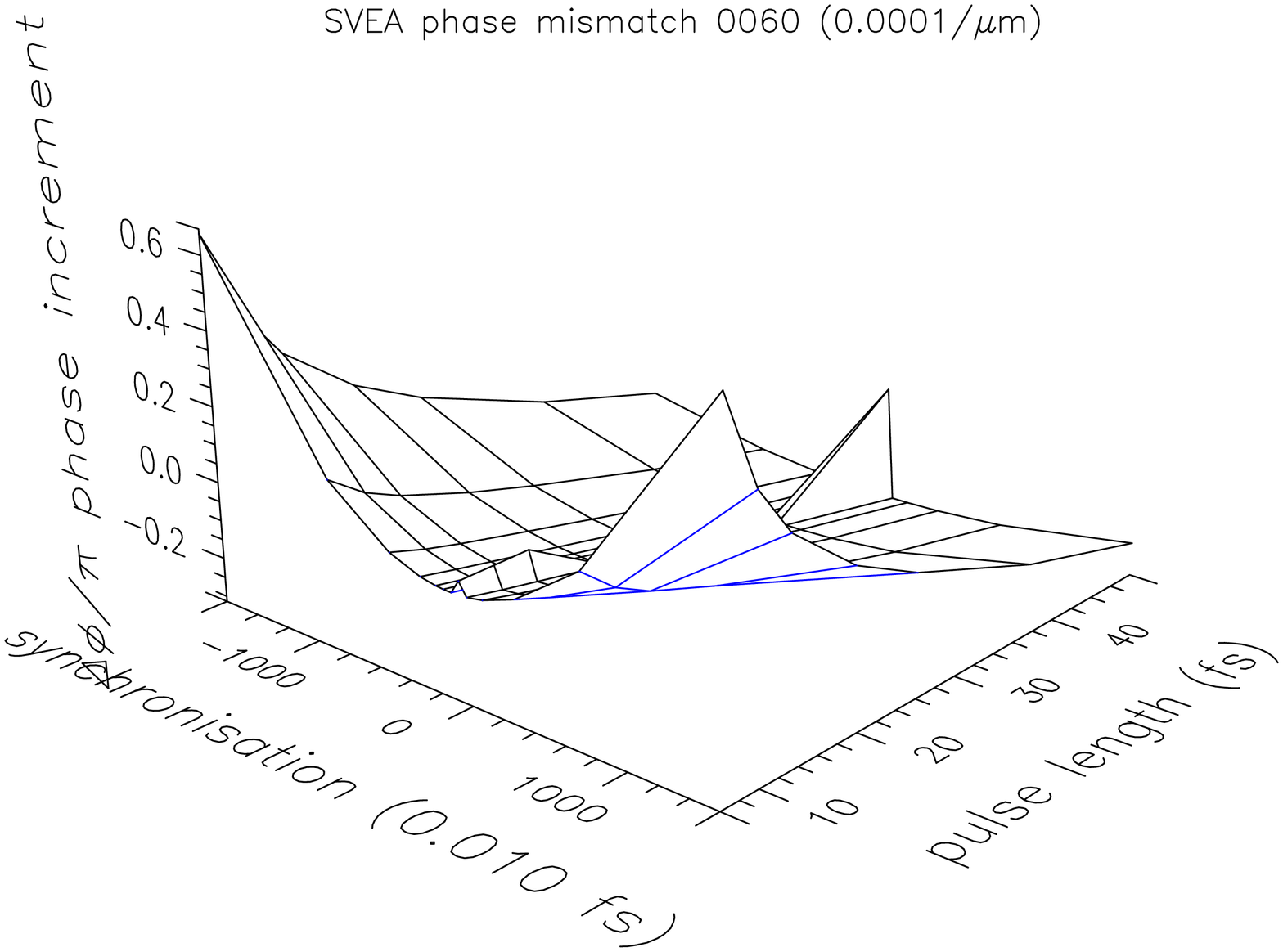}
\includegraphics[width=43mm,angle=-90]{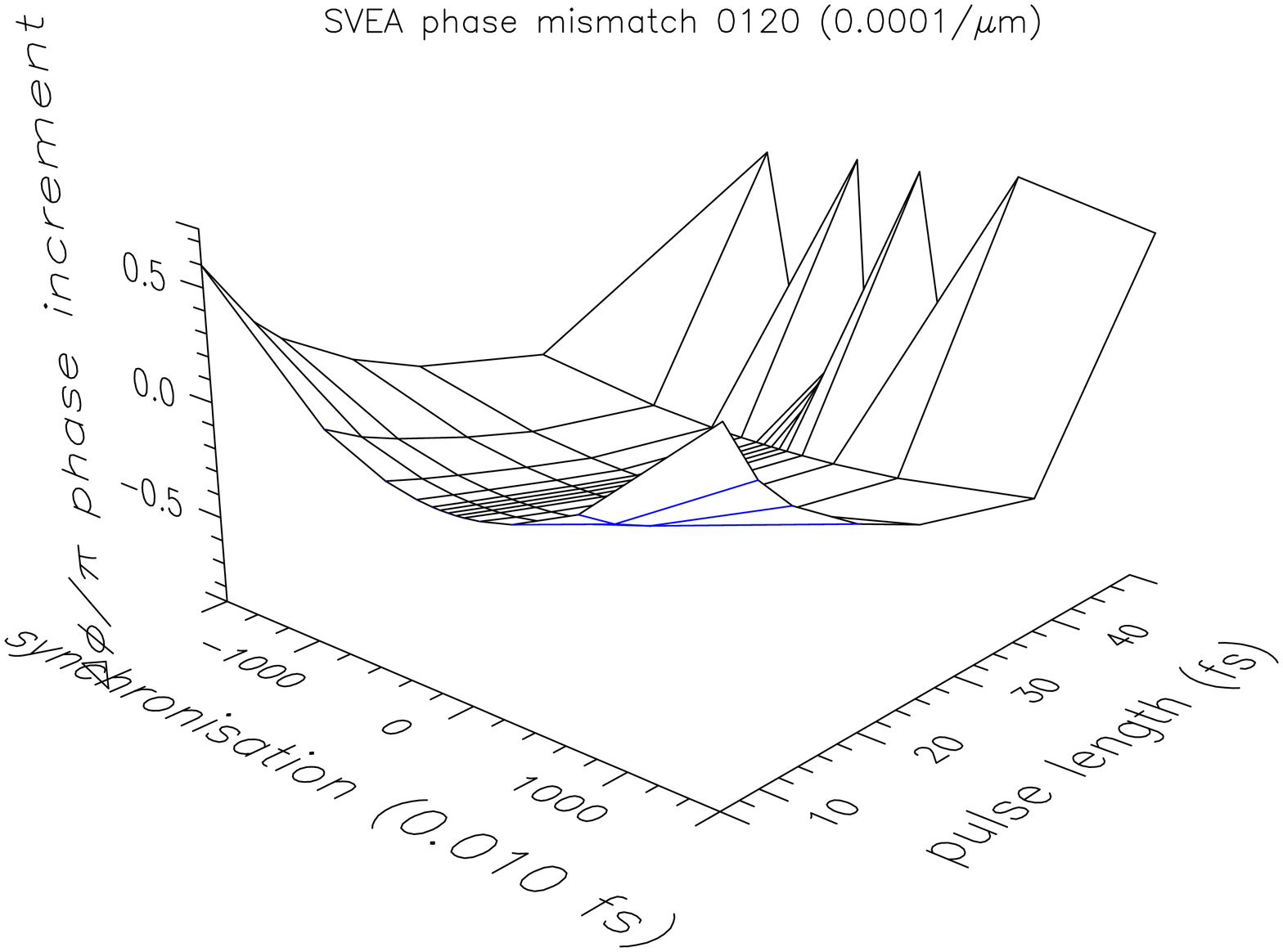}
\includegraphics[width=43mm,angle=-90]{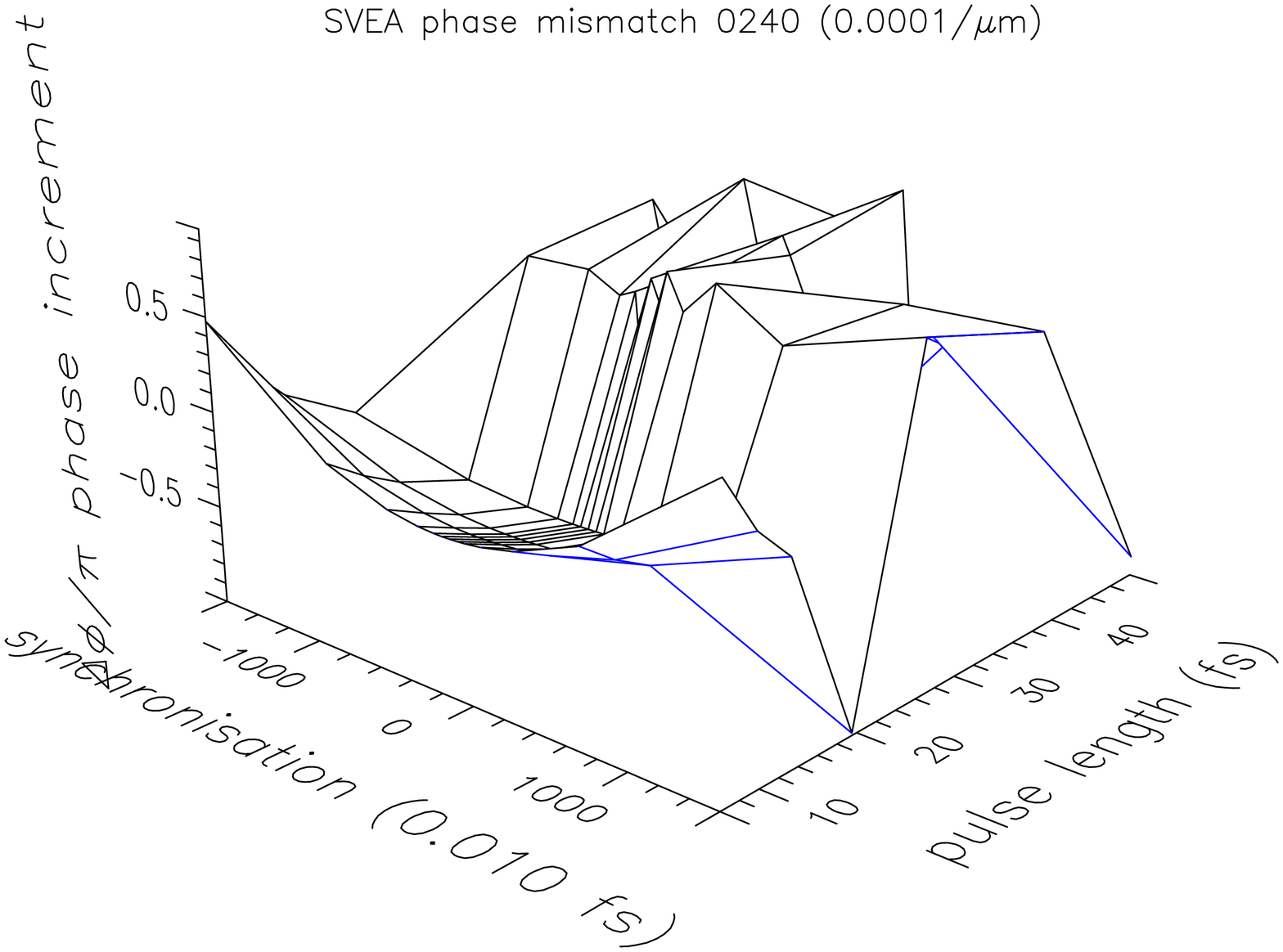}
\caption{ 
\label{Foposyncpp-SVEA}
Scaled OPO: SVEA phase mismatch vs sync 
} 
\end{figure}

\end{subsection}
}
{}


%
%
%
%
%

\ifthenelse{\boolean{BoolOPOfixed}}
{

\begin{subsection}{Fixed Length Crystal}\label{opo-fixed}

Here we present results for a more realistic case of a fixed 
length crystal ($1000\mu$m) and a range of pump pulse 
durations ($6$ -- $96$fs); with all parameters being 
kept constant, notably a fixed pulse energy.  Since the crystal
is relatively long, the high 
dispersion comes into play, ensuring that even for rather 
short pump pulses (e.g. $6$fs), the signal and idler become
relatively broad ($\sim 200$fs) -- thus the role of few-cycle
effects should be relatively small.  However, inspection of 
the envelope profiles for the SVEA and GFEA models still 
show small but noticeable differences, for both perfect and imperfect 
phase matching, even for pump pulses
as long as $96$fs.

\begin{figure}
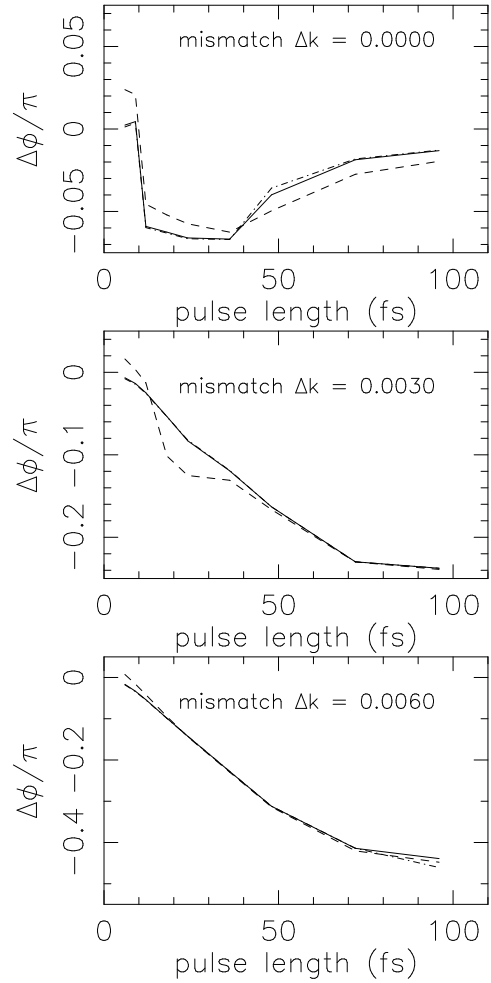

\includegraphics[width=43mm,angle=-90]{Cryst_dphi-1000u-p0000s-0000pp-i0-n9.epsi}
\includegraphics[width=43mm,angle=-90]{Cryst_dphi-1000u-p0000s-0030pp-i0-n9.epsi}
\includegraphics[width=43mm,angle=-90]{Cryst_dphi-1000u-p0000s-0060pp-i0-n9.epsi}
\caption{ 
\label{Fphaseinc1000u}
$1000\mu$m OPO: 
Pass-to-pass phase drifts for a differing pulse lengths but a
fixed crystal length of
1000$\mu$m and equal pulse energies, 
comparing SVEA ({-----}), 
 SEWA ({--~--~--}), 
and GFEA ({--$\cdot$--$\cdot$--$\cdot$})
simulation results.  The differences are taken between the
phase at the peak of the modulus-squared of the envelopes
at the end of one pass of the signal pulse and the next.  Each graph 
is for a different phase mismatch $\Delta k$ in the periodic poling of 
the crystal.
}
\end{figure}

In fig. \ref{Fphaseinc1000u} we see how the inter-pass phase difference
for the signal (and hence idler) for a fixed 1000$\mu$m crystal with 
decreasing pump pulse widths and adjusted time offsets, but fixed pulse energy.
Difference between the SVEA predicted phase drift and the GFEA are 
noteable for perfect phase matching ($\Delta k = 0 $), but when this
is no longer exact, the two predictions rapidly become similar, as now
both a long crystal and imperfect phase matching both act against the 
generation of clear few cycle effects.

\end{subsection}

}


\end{section}



\begin{section}{Conclusions}\label{conclude}

We have presented a new and more complete derivation of how the envelopes of
extremely short optical pulses evolve in nonlinear interactions.  We have
compared the results of our new (GFEA) model to those of the traditional
slowly varying envelope approximation (SVEA) using a scaling procedure to
distinguish specific few-cycle effects from other phenomena caused by changing
pulse duration.  It should be noted that the SVEA becomes inadequate whenever
the envelope changes rapidly within a few carrier periods.  Strictly speaking,
a few-cycle pulse is not required, because a steep edge within a longer pulse
also fulfils the conditions.

The effect of the extra ``few-cycle'' terms in the GFEA evolution
equation is to add a phase distortion to the nonlinear polarization term,
which then imposes itself on the pulse envelopes.  This is demonstrated by
our single-pass optical parametric amplifier NPA model where, whilst the SVEA 
model is insensitive to pulse length, the GFEA theory shows clear changes 
as the pulses get shorter and contain fewer optical cycles.

Further, when we studied the highly sensitive de-amplification case (i.e.
NPD), we saw dramatic differences between the SVEA and GFEA simulations even
outside the few-cycle regime. These arose from the phase distorting effects of
the few-cycle terms in the theory disrupting the exact phase relationships
needed for de-amplification.  While the absolute size of these differences do
depend on the chosen parameters of crystal length, pulse energy, and so on, 
they will always get dramatically larger for shorter pulses, 

On the other hand, the repetitive cycling nature of the optical parametric
oscillator (OPO) produces more complicated and subtle dynamics; small changes
in parameter values can, for instance, cause sudden changes in the steady
state fields.  It is therefore no surprise that comparison of the results
predicted by the different models is less straightforward in the OPO case. 
The new model certainly produces differences in the pulse envelopes as well as
the phases, although the way in which the GFEA tends to the SVEA in the
long-pulse limit has some interesting features.  The two models also predict
different results for the pass-to-pass phase drift of OPO pulses, and this
implies significant differences in the electric field structures.  In both
cases, the carrier wave moves under the envelope from one transit to the next,
but by different amounts.  

\ifthenelse{\boolean{BoolLong}}
{
{It can be useful to regard these ``few-cycle'' effects as adding a 
phase twist to the envelope evolution.  This then shows itself 
most clearly in the pass-to-pass phase drift in the signal and idler
phases (see Fig. \ref{Fphaseincrement}).  Finally, although these 
effects are usually described as ``few-cycle effects'', they are more
accurately described as ``finite pulse length'' effects.  Although 
for the parameter ranges needed to describe typical nonlinear crystals, 
it is only in a few-cycle regime where the effects are easily visible 
in the pulse intensity profiles, these twisting effects on the phase 
structure of the pulse do occur, and could be seen in a many cycle regime.
}
}

\end{section}

{

\end{document}